\documentclass[aps,
		pra,
		reprint,
		twocolumn,
		superscriptaddress,
		shortbibliography,
		nofootinbib,
		floatfix,
		notitlepage
		]{revtex4-1}
\pdfoutput=1

\usepackage[dvipsnames]{xcolor}

\usepackage{hhline}
\usepackage{bm}
\usepackage{amsmath}
\usepackage{amssymb}
\usepackage{amsbsy}
\usepackage{amsfonts}
\usepackage{hyperref}
\usepackage{cleveref}
\usepackage{slashed}
\usepackage{color}
\usepackage{braket}
\usepackage{tikz}
\usepackage{tikz-feynman}
\usetikzlibrary{decorations.pathmorphing,decorations.pathreplacing,decorations.shapes}
\usepackage[utf8]{inputenc}
\usepackage[T1]{fontenc}
\usepackage[normalem]{ulem} 
\usepackage{multirow}

\newcommand{\trm}[1]{\textrm{#1}}
\newcommand{\tsf}[1]{\textsf{#1}}
\newcommand{\ud}{\mathrm{d}}
\newcommand{\HE}{\textsf{HE}}
\newcommand{\Sfi}{\textsf{S}_{\textsf{fi}}}
\newcommand{\Ecr}{E_{\tsf{cr}}}
\newcommand{\eqnref}[1]{Eq. (\ref{#1})}
\newcommand{\figref}[1]{Fig. (\ref{#1})}
\newcommand{\tabref}[1]{Tab. (\ref{#1})}
\newcommand{\secref}[1]{Sec. \ref{#1}}

\newcommand{\ca}{c_{4,1}}
\newcommand{\cb}{c_{4,2}}
\newcommand{\cc}{c_{6,1}}
\newcommand{\cd}{c_{6,2}}
\newcommand{\cgen}{c_{2n,j}}

\newcommand{\clop}{c_{4}^{+}}
\newcommand{\clom}{c_{4}^{-}}
\newcommand{\cnlop}{c_{6}^{+}}
\newcommand{\cnlom}{c_{6}^{-}}

\newcommand{\bragg}{\theta_{\mathsf{B},2n}}

\newcommand{\fnum}{f^{\#}}

\makeatletter
\pgfdeclaredecoration{gluon}{initial}
{
  \state{initial}[
    width=+0pt,
    next state=coil,
    persistent precomputation={
      \pgfmathsetmacro\matchinglength{
        (ceil(\pgfdecoratedinputsegmentlength / \pgfdecorationsegmentlength) - \pgfdecoratedinputsegmentlength / \pgfdecorationsegmentlength) > 0.5
        ? (\pgfdecoratedinputsegmentlength - 2 * \pgfdecorationsegmentaspect * \pgfdecorationsegmentamplitude) / (floor(\pgfdecoratedinputsegmentlength / \pgfdecorationsegmentlength) + 0.499)
        : (\pgfdecoratedinputsegmentlength - 2 * \pgfdecorationsegmentaspect * \pgfdecorationsegmentamplitude) / (ceil(\pgfdecoratedinputsegmentlength / \pgfdecorationsegmentlength) + 0.499)
      }
      \setlength{\pgfdecorationsegmentlength}{\matchinglength pt}
    },
  ]{}
  \state{coil}[
    switch if less than=%
      0.5\pgfdecorationsegmentlength%
      +\pgfdecorationsegmentaspect\pgfdecorationsegmentamplitude%
      +\pgfdecorationsegmentaspect\pgfdecorationsegmentamplitude to last,
    width=+\pgfdecorationsegmentlength,
  ]
  {
    \pgfpathcurveto
    {\pgfpoint@oncoil{0    }{ 0.555}{ 1}}
    {\pgfpoint@oncoil{0.445}{ 1    }{ 2}}
    {\pgfpoint@oncoil{1    }{ 1    }{ 3}}
    \pgfpathcurveto
    {\pgfpoint@oncoil{1.555}{ 1    }{ 4}}
    {\pgfpoint@oncoil{2    }{ 0.555}{ 5}}
    {\pgfpoint@oncoil{2    }{ 0    }{ 6}}
    \pgfpathcurveto
    {\pgfpoint@oncoil{2    }{-0.555}{ 7}}
    {\pgfpoint@oncoil{1.555}{-1    }{ 8}}
    {\pgfpoint@oncoil{1    }{-1    }{ 9}}
    \pgfpathcurveto
    {\pgfpoint@oncoil{0.445}{-1    }{10}}
    {\pgfpoint@oncoil{0    }{-0.555}{11}}
    {\pgfpoint@oncoil{0    }{ 0    }{12}}
  }
  \state{last}[next state=final]
  {
    \pgfpathcurveto
    {\pgfpoint@oncoil{0    }{ 0.555}{1}}
    {\pgfpoint@oncoil{0.445}{ 1    }{2}}
    {\pgfpoint@oncoil{1    }{ 1    }{3}}
    \pgfpathcurveto
    {\pgfpoint@oncoil{1.555}{ 1    }{4}}
    {\pgfpoint@oncoil{2    }{ 0.555}{5}}
    {\pgfpoint@oncoil{2    }{ 0    }{6}}
  }
  \state{final}{}
}
\makeatother

\begin{document}

\title{
Fundamental constants from photon-photon scattering in three-beam collisions
}

\author{A. J.~MacLeod}
\email{alexander.macleod@eli-beams.eu}
\affiliation{ELI Beamlines Facility, The Extreme Light Infrastructure ERIC, Za Radnic\'{i} 835, 25241 Doln\'{i} B\v{r}e\v{z}any, Czech Republic}

\author{B. King}
\affiliation{center for Mathematical Sciences, University of Plymouth, PL4 8AA, UK}

\begin{abstract}
Direct measurement of the elastic scattering of real photons on an electromagnetic field would allow the  fundamental low-energy constants of quantum electrodynamics to be experimentally determined. We show that scenarios involving the collision of three laser beams have several advantages over conventional two-beam scenarios. The kinematics of a three-beam collision allows for a higher signal-to-background ratio in the detection region, without the need for polarimetry and separates out contributions from different orders of photon scattering. A planar configuration of colliding a photon beam from an x-ray free-electron laser with two optical beams is studied in detail. We show that measurements of elastic photon scattering and vacuum birefringence are possible with currently available technology.
\end{abstract}

\maketitle

\section{Introduction \label{sec:Intro}}

Quantum electrodynamics (QED) predicts that the coupling between strong electromagnetic fields mediated by virtual particle-antiparticle pairs imbues the vacuum with nonlinear properties.
This breaks the superposition principle of classical electrodynamics, allowing for the self-interaction of electromagnetic (EM) fields.
Microscopically, this self-interaction corresponds to photon-photon scattering.
Despite being predicted almost a century ago~\cite{Sauter:1931zz,Halpern:1933dya,Heisenberg:1934pza,Heisenberg:1936nmg,Schwinger:1951nm,Toll:1952rq} a direct measurement of on-shell photon-photon scattering and the predicted birefringence of the quantum vacuum have yet to be made, although Delbr\"{u}ck scattering~\cite{Delbruck:1933pla,Schumacher:1975kv} (the scattering of photons in a Coulomb field of a nucleus, i.e. involving off-shell photons) and more recently scattering of quasi-real photons in ultraperipheral collisions of heavy ions in the ATLAS~\cite{ATLAS:2017fur,ATLAS:2019azn} and CMS~\cite{CMS:2018erd} experiments have been observed. 
It has also been reported \cite{Brandenburg:2022tna} that the STAR \cite{STAR:2019wlg} experiment measured an indirect signal of vacuum birefringence in the spectrum of electron-positron pairs created in ultraperipheral heavy-ion collisions. 
The lack of verification of photon-photon scattering with real photons can be  attributed to two obstacles: the small size of the photon-photon scattering cross section and the high background generated when colliding beams of photons.
As laser technology continues to improve and provide photon sources with an ever-higher flux,  experimental verification of real photon-photon scattering seems achievable. 
There have been numerous suggestions for how to increase the signal-to-background ratio in discovery experiments employing intense laser pulses, such as vacuum diffracting a probe beam with an intense pump beam \cite{DiPiazza:2006pr,King:2010nka,King:2010kvw,PhysRevLett.107.053604,Tommasini:2010fb,PhysRevLett.107.073602,King:2012aw,Jin:2022wdz}, vacuum reflection \cite{Gies:2013yxa}, frequency upshifting and downshifting \cite{Lundstrom:2005za,King:2012aw,Gies:2014jia,Aboushelbaya:2019ncg},  suggestions to enhance the vacuum polarization signal by using structured laser pulses \cite{PhysRevLett.107.073602,Aboushelbaya:2019ncg,Formanek:2023mkx,PhysRevA.107.062213}, and strategies to optimize the vacuum polarization signal \cite{Berezin:2024fxt,Valialshchikov:2024svm}. 
Experimental verification of real photon-photon scattering is particularly interesting as it provides the gateway to harnessing the nonlinear response of the vacuum for more exotic applications such as self-focusing \cite{1993ZhETF.103.1996R}, vacuum high harmonic generation \cite{DiPiazza:2005jc,FEDOTOV20071} and vacuum shock waves \cite{Heyl:1998db,Bohl:2015uba} (for reviews on photon-photon scattering and strong-field QED see, e.g.~\cite{Marklund.RevModPhys.78.591.2006,DiPiazza.RevModPhys.84.1177.2012,King:2015tba,Fedotov:2022ely}). 
Furthermore, with construction planned for lasers in the 10-100\,\trm{PW} range \cite{Weber:2017pmh,Shen:2018lbq,danson19,XU2020164553,dipiazza2022multipetawatt}, real photon-photon scattering becomes more easily measurable as technology progresses (see also~\cite{Dinu:2013gaa,Dinu:2014tsa,Bragin:2017yau,Meuren:2014kla,King:2016jnl,Macleod:2023asi}).

In the present paper we highlight the benefits of using a three-beam setup to detect photon scattering and experimentally determine the fundamental low-energy constants of QED. 
The leading-order process is four-photon scattering and by colliding three beams the kinematics of the signal photons can be more directly specified.
Early three-beam experiments \cite{Moulin.Opt.Com.1999} were performed that  bounded the cross section for the process, while investigations in the literature \cite{Lundstrom:2005za,Lundin:2006wu,Gies:2017ezf,King:2018wtn,Ahmadiniaz:2022nrv} and a recent study on signal optimization \cite{Berezin:2024fxt} show how a three-beam setup can produce a change in (i) the polarization, (ii) the momentum, and (iii) the energy of scattering photons, which can all be used to increase the signal-to-background ratio. 
The results we present show that the fundamental low-energy constants of QED, which govern the magnitude of photon-photon scattering and associated phenomena such as vacuum birefringence, could be directly measured with today's technology in a three-beam configuration which collides an x-ray beam with two optical laser pulses.
This complements studies that have bounded combinations of these constants using results from laser-cavity experiments \cite{Fouche:2016qqj} and recent work suggesting colliding two beams to reach the sensitivity required to measure the values of these constants predicted by QED \cite{Karbstein:2022uwf}. 
The most stringent bound on the fundamental low-energy constants has so far been made by the PVLAS laser-cavity experiment \cite{Ejlli:2020yhk}, which by searching for vacuum birefringence has placed a bound on the difference of the two lowest energy constants.

A key advantage of the three-beam setup is that, by an appropriate choice of the collision geometry, signal photons can be scattered out of the probe beam such that in the detector plane there is a spatial separation of signal and background. 
This allows for the photon count to be used as the signal, without requiring extra polarimetry (which, however, for x-ray photons is now sensitive to one in over $10 \times 10^{9}$ \cite{Schulze:2022}) or special modification of the probe beam \cite{Karbstein:2022uwf}. 
If the probe is an x-ray free-electron laser (XFEL), as we consider here, since no extra polarimetry is required there is no requirement on having an especially low bandwidth in the XFEL beam, which in turn allows one to use the self-amplified by spontaneous emission (SASE)~\cite{Kondratenko.Part.Accel.1980,Bonifacio.Opt.Com.1984} mode with a higher photon flux to increase the signal. 
Should a measurement of vacuum birefringence be desired, this can also be facilitated with a three-beam setup. 

A further highlight of the three-beam configuration is that the kinematics also allow for spatially separated signals of $2n$-photon scattering. 
This should be  compared to the two-beam configuration, which heavily suppresses scattering for $n>2$. 
To investigate this point, the present work includes both four-photon and six-photon scattering processes.
This allows us to go beyond previous studies and calculate the sensitivity required to access the fundamental low-energy constants of dimension-8 and dimension-10 operators. 

We begin in \secref{sec:Theory} with some theory background, specify the collision geometry in \secref{sec:Geometry}, and detail the analytical results in \secref{sec:Constants} and numerical results in \secref{sec:Results}.
The paper concludes in \secref{sec:Summary}. Extra technical detail on the beam profiles can be found in the Appendix, which features analysis of the infinite-Rayleigh-length approximation when an angular cut is applied to the scattered field.

\section{Theory Background \label{sec:Theory}}

The nonlinear interaction between photons mediated by virtual electron-positron pairs can be re-cast as an effective field theory with a Lagrangian expressed as a double sum of derivatives and powers of field strengths,
\begin{align}\label{eqn:EffectiveLagrangian}
    \mathcal{L}_{\textsf{eff}} = \sum_{\sigma,n} c^{\textsf{eff}}_{\sigma,n} \partial^{(\sigma)}O^{n}
    \,,
\end{align}
where $\sigma$ counts the number of derivatives and $n$ the number of field strengths in the operator $O$.
The coefficients $c^{\textsf{eff}}_{\sigma,n}$ are fundamental low-energy constants.
When the center-of-momentum energy $\omega_{\star}$ is much lower than the electron rest mass $m_{\mathsf{e}}$ and the field strength is much lower than the Sauter-Schwinger critical field\footnote{Throughout we use units where $\hbar = c = \varepsilon_{0} = 1$.} $\Ecr = m_{\mathsf{e}}^{2}/e\approx1.3\times10^{16}\,\trm{Vcm}^{-1}$ (with $m_{\mathsf{e}}$ and $e>0$ the mass and charge on a positron respectively), the physics can be described by the weak-field expansion of the Heisenberg-Euler Lagrangian~\cite{Heisenberg:1934pza,Heisenberg:1936nmg,Schwinger:1951nm} (see also \cite{Dunne:2004nc}) $\mathcal{L}_{\textsf{eff}} \approx \mathcal{L}_{\HE}$ with
\begin{align}\nonumber
    \mathcal{L}_{\HE}
    =
    &
    \frac{m_{\mathsf{e}}^{4}}{\alpha}
    \sum_{n=2}
    \mathcal{L}_{2n}
    \,,
    \\\nonumber
    \mathcal{L}_{4}
    =
    &
    \frac{\alpha}{360 \pi^{2}}
    \big(
        \ca
        \mathcal{F}^{2}
        +
        \cb
        \mathcal{G}^{2}
    \big)
    \,,
    \\\label{eqn:LagrangianExpand}
    \mathcal{L}_{6}
    =
    &
    \frac{\alpha}{630 \pi^{2}}
    \big(
        \cc
        \mathcal{F}^{3}
        +
        \cd
        \mathcal{F}
        \mathcal{G}^{2}
    \big) 
    \,,
\end{align}
Here $\alpha = e^{2}/4\pi$ is the QED fine-structure constant, $\cgen$ are the fundamental low-energy constants, and throughout we normalize field strengths, e.g., in the Faraday tensor $F^{\mu\nu}$ and its dual $\widetilde{F}^{\mu\nu}$, by the critical field $\Ecr$. The (normalized) EM invariants are then $\mathcal{F} = - F^{\mu\nu} F_{\mu\nu}/4$ and $\mathcal{G} = - \widetilde{F}^{\mu\nu} F_{\mu\nu}/ 4 $ where $\widetilde{F}^{\mu\nu} = \frac{1}{2} \varepsilon^{\mu\nu\alpha\beta} F_{\alpha\beta}$.
The terms $\mathcal{L}_{2n}$ correspond diagrammatically to one-loop $2n$-photon scattering amplitudes,
\begin{align}\label{eqn:ExpansionDiagrams}
    \mathcal{L}_{\HE}
    \sim
    \begin{tikzpicture}[scale=0.3,baseline={([yshift=-0.5ex]current bounding box.center)}]
        \draw[thick] (0.707107,0.707107) edge[color=black,decorate,decoration={gluon, amplitude=2pt, aspect=0}] (1.707107,1.707107);
        \draw[thick] (-0.707107,-0.707107) edge[color=black,decorate,decoration={gluon, amplitude=2pt, aspect=0}] (-1.707107,-1.707107);
        \draw[thick] (-0.707107,0.707107) edge[color=black,decorate,decoration={gluon, amplitude=2pt, aspect=0}] (-1.707107,1.707107);
        \draw[thick] (0.707107,-0.707107) edge[color=black,decorate,decoration={gluon, amplitude=2pt, aspect=0}] (1.707107,-1.707107);
        \draw[thick] (0,0) circle [radius=1];
    \end{tikzpicture}
    ~~
    +
    ~~
    \begin{tikzpicture}[scale=0.3,baseline={([yshift=-0.5ex]current bounding box.center)}]
        \draw[thick] (0.5,0.866025) edge[color=black,decorate,decoration={gluon, amplitude=2pt, aspect=0}] (1.5,1.866025);
        \draw[thick] (-0.5,0.866025) edge[color=black,decorate,decoration={gluon, amplitude=2pt, aspect=0}] (-1.5,1.866025);
        \draw[thick] (0.5,-0.866025) edge[color=black,decorate,decoration={gluon, amplitude=2pt, aspect=0}] (1.5,-1.866025);
        \draw[thick] (-0.5,-0.866025) edge[color=black,decorate,decoration={gluon, amplitude=2pt, aspect=0}] (-1.5,-1.866025);
        \draw[thick] (1,0) edge[color=black,decorate,decoration={gluon, amplitude=2pt, aspect=0}] (2.5,0);
        \draw[thick] (-1,0) edge[color=black,decorate,decoration={gluon, amplitude=2pt, aspect=0}] (-2.5,0);
        \draw[thick] (0,0) circle [radius=1];
    \end{tikzpicture}
    ~~
    +
    ~~
    \dots
    \,.
\end{align}

For the $2\to 2$ process of two photons scattering off each other, 
the total unpolarized photon-photon scattering cross section is~\cite{Berestetskii:1982qgu}
\begin{align}\label{eqn:CrossUnpol}
    \sigma
    =
    &
    \frac{973 \alpha^{4}}{10125 \pi}
    \Big(\frac{\omega_{\star}}{m_{\mathsf{e}}}\Big)^{6}
    \frac{1}{m_{\mathsf{e}}^{2}}
    \quad
    &
    (\omega_{\star} \ll m_{\mathsf{e}})
    \,,
    &
    \nonumber\\
    \sigma
    =
    &
    4.7
    \frac{\alpha^{4}}{\omega_{\star}^{2}}
    \quad
    &
    (\omega_{\star} \gg m_{\mathsf{e}})
    &
    \,.
\end{align}
where $\omega_{\ast}$ is the photon energy in the center-of-momentum frame, i.e., $\omega_{\star}^2=(k_{1}+k_{2})^{2}/4$ for incoming momenta $k_{1}^{\mu}$ and $k_{2}^{\mu}$.
The total cross section peaks at $\omega_{\star}=1.5m_{\mathsf{e}}$~\cite{DeTollis:1965vna} i.e. $\omega_{\star} \approx 0.75~\mathrm{MeV}$, but there are currently no laboratory-based photon sources with a high enough flux at this energy to see real photon-photon scattering in experiment (although recent experiments have bound the cross section \cite{Watt:2024brh}).
If two photons are replaced with couplings to a coherent electromagnetic field, the cross section can be enhanced by the field strength squared. 
By using the strongest terrestrial fields, which are produced by focusing optical beams from high-power laser systems, one can compensate for the small value of the cross section for free photons at these energies $\sigma \sim 10^{-64}~\mathrm{cm}^{2}$.
Another way to enhance photon-photon scattering is to combine optical laser beams with probe photons of a higher energy, such as from a focused XFEL, which gives a center-of-mass energy  $\omega_{\star} \sim 100~\text{eV}$ and the cross section $\sigma \sim 10^{-53}~\mathrm{cm}^{2}$. (For reviews of XFEL theory and facilities see, e.g.,~\cite{Huang.PhysRevSTAB.10.034801.2007,McNeil.Nat.Photonics.2010,Pellegrini.RevModPhys.88.015006.2016,Feng.Nucl.Sci.Tech.2018,huang2021features}.)
The combination of optical and XFEL beams has attracted the interest of several groups investigating photon scattering signals in this setup (see e.g~\cite{Heinzl:2006xc,DiPiazza:2006pr,King:2010kvw,Karbstein:2015xra,Schlenvoigt:2016jrd,Inada:2017lop,Shakeri:2017iph,King:2018wtn,Shen:2018lbq,Huang:2019ojh,Seino:2019wkb,Schmitt:2020ttm,Karbstein:2021ldz,Karbstein:2022uwf,Ahmadiniaz:2022nrv,Formanek:2023mkx,Yu_Xu_Shen_Cowan_Schlenvoigt_2023,Wang:2024syk}).

The collision of the pulses can be formulated as a scattering problem (see, e.g.,~\cite{Galtsov.PLB.1971}), with total amplitude
\begin{align}\label{eqn:Samplitude}
    \Sfi
    =
    -
    i
    \int \ud^{4} x \,\mathcal{L}_{\HE}
    \,.
\end{align}
Here $\mathcal{L}_{\HE}$ is the weak-field expansion of the Heisenberg-Euler Lagrangian \eqnref{eqn:LagrangianExpand} and we  include  terms up to six-photon scattering.
The total EM field can be written as a sum of the colliding and signal fields,
\begin{align}\label{eqn:TotalF}
    F^{\mu\nu}(x)
    =
    F_{\mathsf{x}}^{\mu\nu}(x)
    +
    F_{1}^{\mu\nu}(x)
    +
    F_{2}^{\mu\nu}(x)
    +
    F_{\gamma}^{\mu\nu}(x)
    \,,
\end{align}
where $F_{\mathsf{x}}$ is the x-ray probe field, $F_{1,2}$ are the optical pump fields, and $F_{\gamma}$ is the signal. 
When inserted into the amplitude \eqnref{eqn:Samplitude}, a large number of scattering channels arise. 
However, since the x-ray pulse is produced by an XFEL, which typically has field strengths $F_{\mathsf{x}} \ll F_{1,2}$, for the processes of interest the scattering amplitude can be well-approximated by linearising in the x-ray field and the signal, which is also assumed much smaller than the other fields. 
One can write this linearized amplitude as
\begin{align}\label{eqn:SamplitudeLinear}
    \Sfi
    \approx
    \underbrace{\tsf{S}_{12}
    +
    \tsf{S}_{11}
    +
    \tsf{S}_{22}}_{\trm{four-photon}}
    +
    \underbrace{\tsf{S}_{1122}
    +
    \tsf{S}_{1222}
    +
    \tsf{S}_{1112}}_{\trm{six-photon}}
    \,,
\end{align}
where the subscripts indicate the couplings to the pump fields $F_{1,2}$.

The probe and pump fields are taken to be linearly polarized Gaussian focused pulses in the paraxial approximation, allowing the tensor structure to be separated from the space-time dependence,
\begin{align}\label{eqn:FieldsGaussian}
    F_{j}^{\mu\nu}(x)
    =
    \frac{1}{\omega_{j}}
    \big(
        k^{\mu}_{j} \varepsilon^{\nu}_{j}
        -
        k^{\nu}_{j} \varepsilon^{\mu}_{j}
    \big) 
    E_{j}(x)
    \,,
\end{align}
with wave vectors $k_{j}^{\mu}$, polarization vectors $\varepsilon_{j}^{\mu}$, and field profiles $E_{j}$ for $j \in \{\mathsf{x},1,2\}$. 
(The field profile of the signal photon is similarly defined, but with a plane-wave electric field; see \secref{app:Beams} for details.)
The scattering channels in \eqnref{eqn:SamplitudeLinear} can then be expressed as
\begin{align}\label{eqn:General}
    \tsf{S}_{1^{a}2^{b}}
    =
    T_{ab}
    \int \ud^{4} x~
    E_{\mathsf{x}}(x)
    E_{1}^{a}(x)
    E_{2}^{b}(x)
    E_{\gamma}(x)
    \,,
\end{align}
where $T_{ab}$ depends purely on the tensor structures of each of the fields, and $a$ and $b$ denote the numbers of powers of the fields $E_{1}(x)$ and $E_{2}(x)$ involved in the interaction channel, respectively.
Thus, $\tsf{S}_{12}$, $\tsf{S}_{11}$, and $\tsf{S}_{22}$ in \eqnref{eqn:SamplitudeLinear} give the leading-order (LO) four-photon scattering contribution, while $\tsf{S}_{1122}$, $\tsf{S}_{1112}$, and $\tsf{S}_{1222}$ give the next-to-leading order (NLO) six-photon scattering contribution.

Since the collision is between traveling waves, which have positive- and negative-frequency components, each of the scattering channels can be decomposed into further subchannels with different numbers of photons being absorbed from or emitted into the pump fields.
For example, we can write
\[
E_{j}(x) =\sum_{\delta_{j}}E^{(\delta_{j})}_{j}(x)\,\mbox{e}^{i\delta_{j} k_{j} \cdot x}
\,,
\]
with $\delta_{j} \in \{-1, 1\}$. 
To understand the approximate kinematics one can consider the monochromatic limit, which physically corresponds to infinitely long pulses and zero focusing, where $E^{(\delta_{j})}_{j}(x)$ becomes a constant. 
Then \eqnref{eqn:General} suggests that channel $\tsf{S}_{1^{a}2^{b}}$ supports scattering with momentum conservation of the form
\begin{align}\label{eqn:momCon1}
    \delta_{\mathsf{x}}k_{\tsf{x}} 
    + 
    \underbrace{\delta_{1}k_{1} +\cdots \delta_{1'}k_{1'}}_{a~\trm{terms}} 
    + 
    \underbrace{\delta_{2}k_{2} + \cdots \delta_{2'}k_{2'}}_{b~\trm{terms}} 
    + 
    k_{\gamma}
    =
    0
    \,,
\end{align}
where the $\delta$ terms can take any possible allowed combination and $k_{j'}=k_{j}$ in the monochromatic limit but in general $k_{j'} \neq k_{j}$ in a focused pulse with finite duration. 
Then the most interesting scattering channels are those for which the scattered photon is on-shell, i.e. $k_{\gamma}\cdot k_{\gamma}=0$, and hence can reach the detector.

The number of scattered photons $\tsf{N}_{\gamma}$ can be calculated in the usual way by modulus squaring the amplitude and integrating over the scattered photon momentum. 
This can be reformulated as a differential number of photons:
\begin{align}\label{eqn:NumberGamma}
    \frac{\ud^{3} \tsf{N}_{\gamma}^{\parallel,\perp}}{\ud \omega_{\gamma} \ud \theta_{\gamma} \ud \phi_{\gamma}}
    =
    &
    V^{2}\frac{\omega_{\gamma}^{2} \sin\theta_{\gamma}}{(2\pi)^{3}}
|\Sfi(\omega_{\gamma},\theta_{\gamma},\phi_{\gamma},\varepsilon_{\gamma}^{\parallel,\perp})|^{2}
    \,,
\end{align}
where $V$ is a volumetric normalization factor and the amplitude $\Sfi$ is now written explicitly as a function of parameters of the scattered photon, with frequency $\omega_{\gamma} = k^{0}_{\gamma}$ and polarization basis $\varepsilon_{\gamma}^{\parallel,\perp}$ to be later defined.
The solid angle coordinates $\theta_{\gamma}$ and$\phi_{\gamma}$ are defined with respect to the x-ray propagation direction (corresponding to $\theta_{\gamma}=0$). 
The total number of signal photons is then
\begin{align}\label{eqn:NumberTotal}
    \tsf{N}_{\gamma} = \tsf{N}_{\gamma}^{\parallel} + \tsf{N}_{\gamma}^{\perp}
    \,.
\end{align}

\section{Collision geometry and kinematics \label{sec:Geometry}}

\begin{figure}[t!!]
    \includegraphics[width=0.4\textwidth,trim={0.0cm 0.0cm 0.0cm 0.0cm},clip=true]{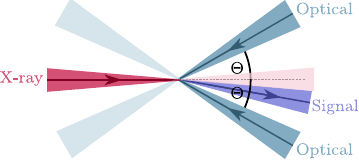}
    \caption{\label{fig:Schematic} Schematic of the planar three-beam collision. An x-ray beam collides with two optical beams which are at angles $\Theta$ and $-\Theta$ to the counterpropagating direction. (Faint colours show the trajectory of the beams after the collision.)}
\end{figure}

We consider a planar three-beam configuration where two optical pump fields, which are chosen to have the same form, collide with an XFEL probe field at an angle $\Theta$ from head-on and with each other at an angle $2\Theta$ as depicted in \figref{fig:Schematic}. 
The x-ray propagates down the $z$ axis and the wave vectors are
\begin{align}\label{eqn:SpecificMomenta}
    k_{\tsf{x}}^{\mu} = & \omega_{\tsf{x}} (1,0,0,1)
    \,, 
    \nonumber\\
    k_{1}^{\mu}(\Theta)
    =
    &
    \omega_{0} (1, \sin\Theta, 0, -\cos\Theta)
    =
    k_{2}^{\mu}(-\Theta)
    \,,
    \nonumber \\ 
    k_{\gamma}^{\mu}
    =
    &
    \omega_{\gamma}    (1,\sin\theta_{\gamma}\cos\phi_{\gamma},\sin\theta_{\gamma}\sin\phi_{\gamma},\cos\theta_{\gamma})
    \,.
\end{align}
For each wave vector one can form a corresponding transverse polarization basis $\{\alpha_{i}^{\mu},\beta_{i}^{\mu}\}$ ($i = \{\mathsf{x},1,2\}$), where $k_{i} \cdot \alpha_{i} = k_{i} \cdot \beta_{i} = \alpha_{i} \cdot \beta_{i} = 0$ and $\alpha_{i} \cdot \alpha_{i} = \beta_{i} \cdot \beta_{i} = -1$.
The basis polarization vectors for the XFEL beam are
\begin{align}\label{eqn:XFELbasis}
    \alpha_{\tsf{x}}^{\mu}
    =
    &
    (0,1,0,0)
    \,,
    \quad
    &
    \beta_{\tsf{x}}^{\mu}
    =
    &
    (0,0,1,0)
    \,,
\end{align}
such that a general linear polarization state is
\begin{align}\label{eqn:XFELparallel}
    \varepsilon_{\tsf{x}}^{\mu}(\psi_{\tsf{x}}) 
    =
    &
    \left(\cos\psi_{\tsf{x}}\right)
    \alpha_{\tsf{x}}^{\mu}
    +
    \left(\sin\psi_{\tsf{x}}\right)
    \beta_{\tsf{x}}^{\mu}
    \,.
\end{align}
The corresponding orthonormal polarization state is:
\begin{align}\label{eqn:XFELperpendicular}
    \bar{\varepsilon}_{\tsf{x}}^{\mu}(\psi_{\tsf{x}}) 
    =
    \left(\sin\psi_{\tsf{x}}\right)
    \alpha_{\tsf{x}}^{\mu}
    -
    \left(\cos\psi_{\tsf{x}}\right)
    \beta_{\tsf{x}}^{\mu}
    \,.
\end{align}
The angle $\psi_{\tsf{x}}$ defines the polarization plane of the x-ray beam and will be important in maximizing, e.g., the number of polarization-flipped photons. 
The basis polarization states for the optical beams are, for $k_{1}^{\mu}(\Theta)$,
\begin{align}\label{eqn:Optical1basis}
    \alpha_{1}^{\mu}
    =
    &
    (0,\cos\Theta,0,\sin\Theta)
    \,,
    \quad
    &
    \beta_{1}^{\mu}
    =
    &
    (0,0,1,0)
    \,,
\end{align}
and for $k_{2}^{\mu}(\Theta)$,
\begin{align}\label{eqn:Optical2basis}
    \alpha_{2}^{\mu}
    =
    &
    (0,\cos\Theta,0,-\sin\Theta)
    \,,
    \quad
    &
    \beta_{2}^{\mu}
    =
    &
    (0,0,1,0)
    \,.
\end{align}
Note that $\beta_{\tsf{x}}^{\mu} = \beta_{1}^{\mu} = \beta_{2}^{\mu}$ as the planar geometry ensures there is a common orthogonal direction normal to the interaction plane.
As with the XFEL beam, it is convenient to parametrize the general linear polarization states of the optical beams through angles $\psi_{1}$ and $\psi_{2}$,
\begin{align}\label{eqn:Optical1pol}
    \varepsilon_{j}^{\mu}(\psi_{j})
    =
    \left(\cos\psi_{j}\right)
    \alpha_{j}^{\mu}
    +
    \left(\sin\psi_{j}\right)
    \beta_{j}^{\mu}
    \,,
\end{align}
for $j\in\{1,2\}$. 
The polarization vector of the signal photon, $\varepsilon_{\gamma}^{\mu}$ is defined as,
\begin{align}\label{eqn:SignalPol}
    \varepsilon_{\gamma}^{\mu}
    =
    e^{\mu}
    -
    \frac{
        e \cdot k_{\gamma}
    }{k \cdot k_{\gamma}}
    k^{\mu}
    \,,
\end{align}
where $e^{\mu}$ is a polarization direction and $k^{\mu}$ is an arbitrary light-like reference vector.
With this definition the signal photon satisfies $k_{\gamma} \cdot \varepsilon_{\gamma} = 0$ by construction.
Without loss of generality we choose the light-like reference vector as $k^{\mu} = k_{1}^{\mu}$.
We will be interested in exploring vacuum birefringence effects, which are microscopically described by polarization flip of the probe photon from the initial polarization state to the orthonormal state. 
Thus, parallel polarized signal photons are defined as non-flipped photons with $e^{\mu} = \varepsilon_{\tsf{x}}^{\mu}$,
\begin{align}\label{eqn:Signalparallel}
    \varepsilon_{\gamma}^{\parallel \mu}(\psi_{\tsf{x}})
    \equiv
    &
    \varepsilon_{\tsf{x}}^{\mu}(\psi_{\tsf{x}})
    -
    \frac{
        k_{\gamma} \cdot \varepsilon_{\tsf{x}}(\psi_{\tsf{x}}) 
    }{k_{1} \cdot k_{\gamma}}
    k_{1}^{\mu}
    \,,
\end{align}
and perpendicular polarized signal photons are those where the polarization has flipped $e^{\mu} = \bar{\varepsilon}_{\tsf{x}}^{\mu}$,
\begin{align}\label{eqn:Signalperpendicular}
    \varepsilon_{\gamma}^{\perp \mu}(\psi_{\tsf{x}})
    \equiv
    &
    \bar{\varepsilon}_{\tsf{x}}^{\mu}(\psi_{\tsf{x}})
    -
    \frac{
        k_{\gamma} \cdot \bar{\varepsilon}_{\tsf{x}}(\psi_{\tsf{x}}) 
    }{k_{1} \cdot k_{\gamma}}
    k_{1}^{\mu}
    \,.
\end{align}

The scattering channels that are of most interest are the elastic channels that allow for an on-shell photon to be scattered out of the main background of the XFEL beam. 
From the general momentum conservation equation \eqnref{eqn:momCon1}, these channels have $\delta_{1}=-\delta_{2}$ for the LO process of four-photon scattering and also $\delta_{1'}=-\delta_{2'}$ for the NLO process of six-photon scattering.
Using the specific setup in \eqnref{eqn:SpecificMomenta}, this corresponds to a photon scattered via $2n$-photon scattering with a momentum
\begin{align}\label{eqn:EnergyMomentumKick2}
    k_{\gamma}^{\mu}
    =
    k_{\tsf{x}}^{\mu}
    \pm
    (n - 1)
    \big[
        k_{1}^{\mu}(\Theta)
        -
        k_{2}^{\mu}(\Theta)
    \big] 
    \,.
\end{align}    
Although $k_{\gamma}\cdot k_{\gamma} \neq 0$, the beams are also not monochromatic. 
If one allows for a finite bandwidth by replacing $k_{\gamma} \to k_{\gamma} + \Delta$, where the bandwidth $\Delta$ is the momentum difference due to taking momenta from three pulses with finite bandwidths, then it follows that the condition $k_{\gamma}\cdot k_{\gamma}=0$ implies that
\[
k_{\tsf{x}}\cdot \Delta - (n-1)^{2} k_{1}\cdot k_{2} \pm (n-1)\, \Delta \cdot \left(k_{1}-k_{2}\right) + \frac{1}{2}\,\Delta^{2} = 0
\,,
\]
where $k_{1,2} \equiv k_{1,2}(\Theta)$.
Since the x-ray frequency is approximately $O(10^{5})$ times larger than the optical frequency, the requirement on the bandwidth $\Delta$ to satisfy the on-shell condition for the scattered photon is easily fulfilled.
To parametrize the small scattering angles of the x-ray photons it will also be useful to define the angles $\theta_{x} = \sin\theta_{\gamma}\cos\phi_{\gamma}$ and $\theta_{y} = \sin\theta_{\gamma}\sin\phi_{\gamma}$ such that the signal photon is parametrized by
\begin{align}\label{eqn:Signalwave vector2}
    k_{\gamma}^{\mu}
    =
    \omega_{\gamma}
    \big(
        1,\theta_{x},\theta_{y},\sqrt{1 - \theta_{x}^{2}-\theta_{y}^{2}}
    \big)
    \,,
\end{align}
with the requirement that $\theta_{x,y}\ll 1$.

\begin{figure}[t!!]
    \includegraphics[width=0.55\textwidth,trim={1.0cm 0.0cm 0.0cm 0.0cm},clip=true]{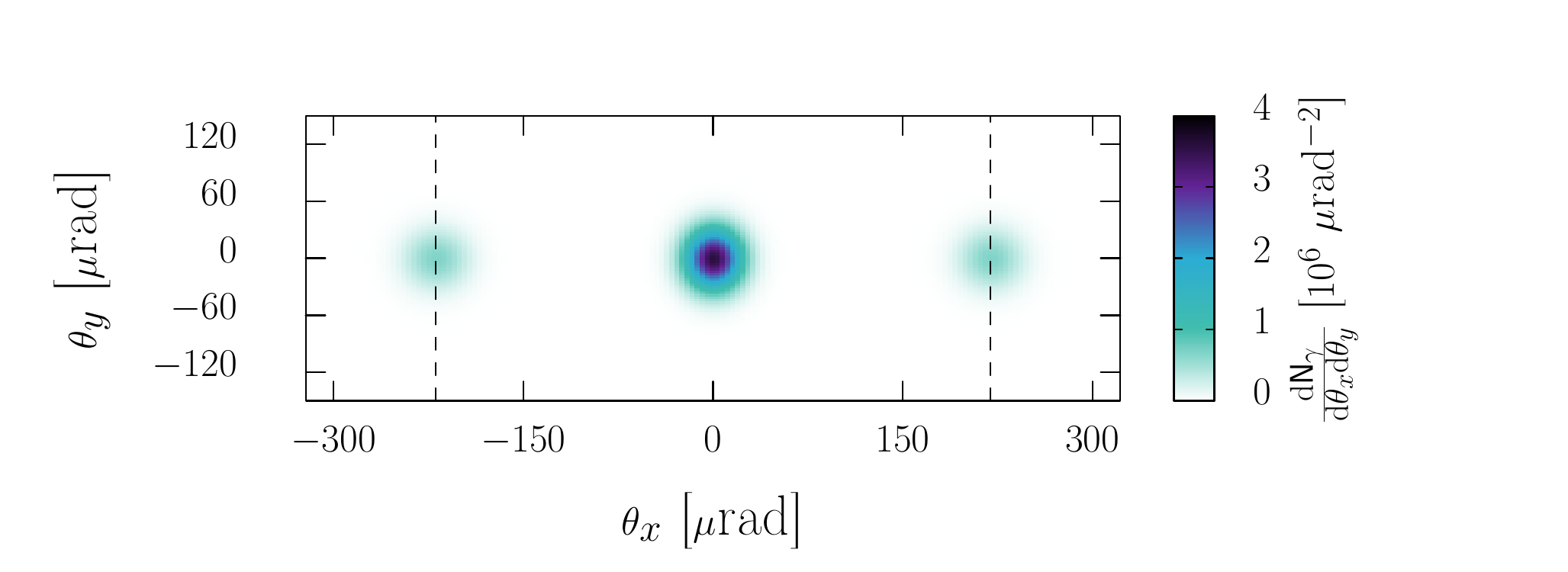}
    \caption{\label{fig:AngleCurrent} 
        Differential number of signal photons $\frac{\ud \mathsf{N}_{\gamma}}{\ud \theta_{x} \ud \theta_{y}}$ in three-beam configuration using EuXFEL SASE parameters (see \tabref{tab:Parameters} and the discussion of \secref{sec:Results} for more details) with $\fnum = 1$ optical focusing and an x-ray beam waist $w_{\tsf{x}} = 4~\text{\textmu{}m}$.
        The collision angle is $\Theta = 45^{\circ}$ and x-ray and optical beams have a relative polarization of $\Delta\psi = \pi/2$.
        The angles $\theta_{x}$ and $\theta_{y}$ are the scattering angles of the photons inside and out the interaction plane, respectively [see \eqnref{eqn:Signalwave vector2}].
        Dashed vertical lines are at Bragg peak locations [\eqnref{eqn:BraggPeak}].
    }
\end{figure}

\begin{figure}[t!!]
    \includegraphics[width=0.55\textwidth,trim={2.0cm 0.0cm 0.0cm 0.0cm},clip=true]{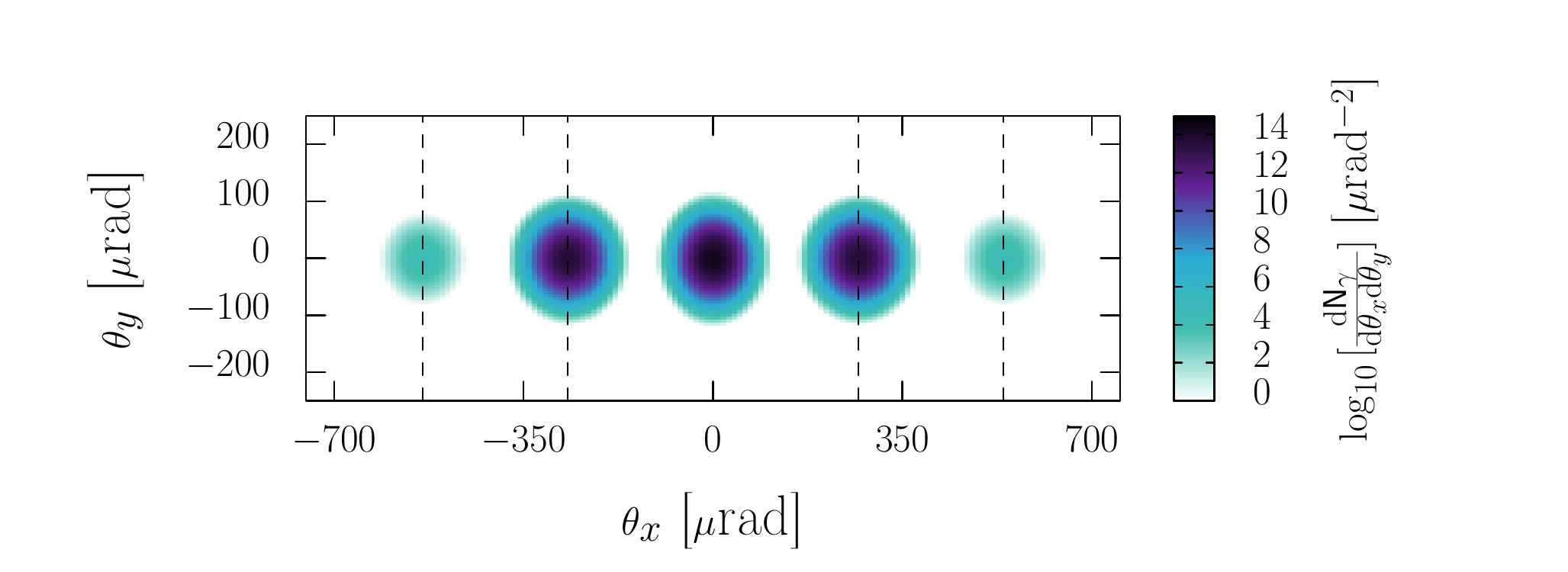}
    \caption{\label{fig:AngleFuture} 
        Logarithmic differential number of signal photons $\log_{10}\left[\ud \mathsf{N}_{\gamma}/\ud \theta_{x} \ud \theta_{y}\right]$ in three-beam configuration using future SASE parameters (see \tabref{tab:Parameters} and the discussion of \secref{sec:Results} for more details) with $\fnum = 2$ optical focusing and an x-ray beam waist $w_{\tsf{x}} = 2~\text{\textmu{}m}$.
        The collision angle is $\Theta = 63^{\circ}$ and x-ray and optical beams have polarization $\psi_{\mathsf{x}} = \psi_{0} = 0$.
        Dashed vertical lines are at Bragg peak locations [\eqnref{eqn:BraggPeak}].
    }
\end{figure}

The scattering channels $\tsf{S}_{11}$ and $\tsf{S}_{22}$ in \eqnref{eqn:SamplitudeLinear} correspond to the case that a photon is absorbed and emitted to the same beam, which by energy momentum conservation gives $k_{\gamma}^{\mu} \approx k_{\tsf{x}}^{\mu}$ (where the $\approx$ sign indicates some allowance for a small spread of momenta due to finite bandwidth effects as explained in the preceding paragraph). 
For these channels the signal photons receive no lateral momentum kick and peak around $\theta_{\gamma} \approx 0$ along the propagation direction of the probe XFEL beam.
Instead, the main interest in the three-beam configuration is in the $\tsf{S}_{12}$ and $\tsf{S}_{1122}$ channels of \eqnref{eqn:SamplitudeLinear}, which includes absorption of photons from one optical beam and emission to the other. 
This produces Bragg side peaks outside the emission cone of the XFEL probe beam, centered around $\theta_{x} \approx \bragg \ll 1$, where
\begin{align}\label{eqn:BraggPeak}
    \bragg
    =
    2 (n - 1)
    \frac{\omega_{0}}{\omega_{\tsf{x}}}
    \sin\Theta
    \,.
\end{align}
An example of the distribution of scattered photons for currently attainable parameters is shown in \figref{fig:AngleCurrent}. 
Only two side peaks are visible, corresponding to four-photon scattering and $n=2$ in \eqnref{eqn:EnergyMomentumKick2}. 
At the level of the probability, the coupling to each optical field scales as $E_{j}^{2} \ll 1$ (recall the fields are normalized by the Schwinger critical field strength) and since six-photon scattering has two extra vertices coupled to optical fields compared with four-photon scattering, the six-photon scattering signal is very weak.
By momentum conservation in \eqnref{eqn:EnergyMomentumKick2} with $n=3$ there are two further Bragg side peaks and these are demonstrated in \figref{fig:AngleFuture} on a logarithmic plot with future parameters for the optical beam.
(See \secref{sec:Results} below for more details on the parameters used.)

\section{Determination of fundamental low-energy constants \label{sec:Constants}}

First we present analytical results that link the number of polarized and unpolarized scattered photons to the fundamental low-energy constants in \eqnref{eqn:LagrangianExpand}.
To simplify the analysis we set the polarization plane of the optical beams to be equal, $\psi_{1}=\psi_{2}=\psi_{0}$. 
Then only two polarization angles remain: $\psi_{\tsf{x}}$ for the x-ray polarization plane and $\psi_{0}$ for the optical polarization plane.
We can obtain more manageable expressions for the number of signal photons by considering an expansion of the amplitudes in \eqnref{eqn:NumberGamma} for $\theta_{x} \ll 1$, or equivalently by taking $\theta_{\gamma} \ll 1$ (see \eqnref{eqn:Signalwave vector2}).
In fact, expanding only the tensor structure, i.e. $T_{ab}$ in \eqnref{eqn:General}, one finds that the parallel and perpendicular amplitudes can be expressed as
\begin{align}\label{eqn:AmplitudeParallelApprox}
    \lim\limits_{\theta_{\gamma} \to 0} \Sfi(\omega_{\gamma},\theta_{\gamma},\phi_{\gamma},\varepsilon_{\gamma}^{\parallel,\perp}) 
    \approx
    &
    g_{\parallel,\perp}
    G
    +
    h_{\parallel,\perp}
    H
    \,,
\end{align}
where $g_{\parallel}$, $g_{\perp}$, $h_{\parallel}$, and $h_{\perp}$ depend only on the $\cgen$ and the initial beam polarizations, $\psi_{\tsf{x}}$ and $\psi_{0}$.
Explicitly,
\begin{align}\nonumber
    g_{\parallel}
    =
    &
    \clop 
    +
    \clom
    \cos\left[2 (\psi_{\tsf{x}} - \psi_{0})\right]
    \,,
    \\\nonumber
    g_{\perp}
    =
    &
    -
    \clom
    \sin\left[2 (\psi_{\tsf{x}} - \psi_{0})\right]
    \,,
    \\\nonumber
    h_{\parallel}
    =
    &
    (2 \cnlop + \cnlom)
    \cos(2 \psi_{\tsf{x}})
    +
    3 \cnlom
    \cos\left[2 (\psi_{\tsf{x}} - 2 \psi_{0})\right]
    \nonumber \\ \nonumber
    & 
    +
    2
    (2 \cnlop + \cnlom)
    \cos(2 \psi_{0})
    \,,
    \\\label{eqn:hper}
    h_{\perp}
    =
    &
    (2 \cnlop + \cnlom)       
    \sin(2 \psi_{\tsf{x}})
    +
    3 \cnlom
    \sin\left[2 (\psi_{\tsf{x}} - 2 \psi_{0})\right]
    \,,
\end{align}
where $\clop=\ca+\cb$, $\clom=\ca-\cb$ and $\cnlop=\cc + \cd$, $\cnlom=\cc - \cd$.
The functions $G$ and $H$ contain all of the space-time dependent information 
\begin{align}
    G
    =
    &
    4
    \cos^{4}\Big(\frac{\Theta}{2}\Big) 
   \int \ud^{4} x \,
    E_{\mathsf{x}}(x)
    E_{\gamma}(x)
    \big[
        E_{1}(x)
        +
        E_{2}(x)
    \big]^{2}
    \,,
     \nonumber\\\label{eqn:H}
    H
    =
    &
    -
    16
    \sin^{2}\Big(\frac{\Theta}{2}\Big)
    \cos^{6}\Big(\frac{\Theta}{2}\Big) 
    \\ \nonumber
    & 
    \times 
    \int \ud^{4} x\,
    E_{\mathsf{x}}(x)
    E_{1}(x)
    E_{2}(x)
    E_{\gamma}(x)
    \left[
        E_{1}(x)
        +
        E_{2}(x)
    \right]^{2}
    \,,
\end{align}
and, in general, $G \gg H$, since the electric fields are normalized by the critical field.
We note here the different dependence of the integral prefactors in $G$ and $H$ on the collision angle $\Theta$.
On the interval $0 < \Theta < \pi/2$, the prefactor of $G$ is a monotonically decreasing function of $\Theta$, with a maximum at $\Theta = 0$.
Conversely, the prefactor of $H$ peaks around $\Theta = 60^{\circ}$ and vanishes for $\Theta = 0$.
This emphasizes the fact that the NLO contribution is heavily suppressed in a head-on two-beam collision and indicates a preferred geometry for the three-beam NLO interaction.
Later numerical investigation will confirm the NLO signal peaks at $\Theta \approx 60^{\circ}$.

Consider now the (differential) number of photons scattered into each of the polarization modes [\eqnref{eqn:NumberGamma}].
With the tensor structure expanded as before, to leading order in $\theta_{\gamma} \ll 1$, these will take the form,
\begin{align}\label{eqn:NumberParApprox}
    \frac{\ud^{3} \tsf{N}_{\gamma}^{\parallel,\perp}}{\ud \omega_{\gamma} \ud \theta_{\gamma} \ud \phi_{\gamma}}
    \approx
    &     
    \frac{\theta_{\gamma}}{(2\pi)^{3}}
    \left[
        g_{\parallel,\perp}^{2}
        |G|^{2}
        +
        h_{\parallel,\perp}^{2}
        |H|^{2} \right. \nonumber \\
        & \left. 
        \quad +
        2
        |g_{\parallel,\perp} h_{\parallel,\perp}|\,
        \mathrm{Re}(H G^{\dagger})
     \right] 
     \,.
\end{align} 
These terms contain a dominant contribution from the LO four-photon interaction, proportional to $|G|^{2}$, a term which depends only on the NLO six-photon interaction, proportional to $|H|^{2}$, and an interference term which mixes the LO and NLO contributions, proportional to $\mathrm{Re}(HG^{\dagger})$.
Since, in general, $G \gg H$, any kinematic region in which the contributions from the LO and NLO diagrams overlap will be dominated by the LO process. We first concentrate on determination of the fundamental low-energy constants of LO scattering.

\subsubsection*{Fundamental constants of four-photon scattering} 

Due to the additional suppression of NLO scattering by powers of the critical field, in detector regions where the contributions from LO and NLO scattering overlap the NLO terms in \eqnref{eqn:NumberParApprox} can be neglected, giving the simple expressions,
\begin{align}
    \tsf{N}_{\gamma}^{\parallel}
    \approx
    &
    \left[
        \ca + \cb 
        + 
        (\ca - \cb) \cos(2 \Delta\psi)
    \right]^{2}
    \mathcal{I}_{G}
    \,,
    \nonumber\\
    \tsf{N}_{\gamma}^{\perp}
    \approx
    &
    \left[
        (\ca - \cb)
        \sin(2 \Delta\psi)
    \right]^{2}
    \mathcal{I}_{G}
    \,,
    \nonumber\\
    \label{eqn:NumberTotalApproxLO}
    \tsf{N}_{\gamma}
    \approx
    &
    2\left[
        \ca^2 + \cb^2 + (\ca^2-\cb^2)
        \cos(2 \Delta\psi)
    \right]
    \mathcal{I}_{G}
    \,.
\end{align}
The function $\mathcal{I}_{G}$ contains all of the space-time structure and integrations which will strongly depend on the experimental pulse parameters and shot-to-shot fluctuations.
The field-independent prefactors of $\mathcal{I}_{G}$ coincide with known scaling laws found for the case of a two-beam head-on collision (see, for example,~\cite{Karbstein:2022uwf}), and depend only on the four-photon low-energy constants~\cite{Ritus:1975pcc},
\begin{align}\label{eqn:Coefficients12}
    \ca
    =
    &
    4
    \bigg(
        1
        +
        \frac{40}{9}
        \frac{\alpha}{\pi}
    \bigg)
    \,,
    \quad
    &
    \cb
    =
    &
    7
    \bigg(
        1
        +
        \frac{1315}{252}
        \frac{\alpha}{\pi}
    \bigg)
    \,,
\end{align}
and the relative polarization of the x-ray and optical beams $\Delta\psi = \psi_{\tsf{x}} - \psi_{0}$.
When $\Delta\psi = \pi/4$, $\tsf{N}_{\gamma}^{\perp}$ takes its maximum value of $\tsf{N}_{\gamma}^{\perp} \approx (\ca-\cb)^2 \mathcal{I}_{G}$; when $\Delta\psi = \pi/2$,  $\tsf{N}_{\gamma}^{\parallel}$ and $\tsf{N}_{\gamma}$ take their maximum values $\tsf{N}_{\gamma}^{\parallel} \approx \tsf{N}_{\gamma} \approx \left(2 \cb\right)^{2} \mathcal{I}_{G}$.
There are two different types of measurement that can be performed to determine the fundamental low-energy constants $\ca$ and $\cb$.
  
(i) \emph{Polarization-sensitive measurement of the number of photons at a fixed angle of polarization of the XFEL beam}. 
In this case, $\Delta \psi$ is fixed and by forming the ratio of $\tsf{N}_{\gamma}^{\parallel}$ to $\tsf{N}_{\gamma}^{\perp}$ we see
\begin{align}\label{eqn:RatioLO}
    \frac{\tsf{N}_{\gamma}^{\perp}}{\tsf{N}_{\gamma}^{\parallel}}
    \approx
    \bigg(
        \frac{(\ca - \cb) \sin(2 \Delta\psi)}{\ca + \cb + (\ca-\cb) \cos(2 \Delta\psi)}
    \bigg)^{2}
    \,,
\end{align}
which is independent of $\mathcal{I}_{G}$ and not affected by experimental factors such as the field-strengths, space-time overlap, or collision angle $\Theta$ of the three beams.

(ii) \emph{Polarization-insensitive measurement of just the number of scattered photons $\mathsf{N}_{\gamma}$ at two different angles of polarization of the XFEL beam}. 
In this case, the two angles of polarization correspond to $\Delta\psi$ with a photon count $\tsf{N}_{\gamma}(\Delta\psi)$ and $\Delta\psi^{\prime}$ with a photon count $\tsf{N}_{\gamma}(\Delta\psi^{\prime})$, with the condition $\Delta\psi^{\prime} \ne \Delta\psi$. The ratio of these two measurements would also be independent of $\mathcal{I}_{G}$,
\begin{align}\label{eqn:NRatioLO}
    \frac{\tsf{N}_{\gamma}(\Delta\psi)}{\tsf{N}_{\gamma}(\Delta\psi^{\prime})}
    =
    \frac{
        \ca^{2}
        +
        \cb^{2}
        +
        (\ca^{2} - \cb^{2})
        \cos(2 \Delta\psi)
    }{
        \ca^{2}
        +
        \cb^{2}
        +
        (\ca^{2} - \cb^{2})
        \cos(2 \Delta\psi^{\prime})
    }
    \,.
\end{align}

These two routes to determining the fundamental low-energy constants $\ca$ and $\cb$ have been outlined in the literature for the two-beam case \cite{Mosman:2021vua}. 
However, a clear advantage of the three-beam setup is that the photons are scattered into side peaks and hence are spatially separated from the large x-ray background from the probe XFEL beam.
This may prove to be more attractive experimentally than modifying the XFEL beam with advanced techniques such as the shadow technique~\cite{Karbstein:2022uwf}.

A further advantage is the choice of the XFEL beam mode. 
X-ray polarimetry, for example, using quasi-channel-cut crystals, can set a strict bound on the allowable bandwidth $\Delta\omega_{\tsf{x}}$ of the x-ray pulse to the order of $100~\mathrm{meV}$~\cite{Ahmadiniaz:2024xob}. 
If a measurement requires a high level of x-ray polarimetry this limits the mode of the XFEL beam to use a low-bandwidth option such as self-seeding.
However, if polarization-insensitive measurements can be used, such as suggested by the three-beam configuration, there is no strict requirement on the x-ray bandwidth and one can use alternative x-ray modes that offer larger numbers of photons per pulse, such as SASE~\cite{Kondratenko.Part.Accel.1980,Bonifacio.Opt.Com.1984}.
Since the total number of signal photons is directly proportional to the number of probe x-ray photons using, e.g., the SASE mode can lead to a significant increase in the signal versus a polarization-sensitive measurement.

\begin{figure}[t!!]
    \includegraphics[width=0.45\textwidth,trim={0.0cm 0.0cm 0.0cm 0.0cm},clip=true]{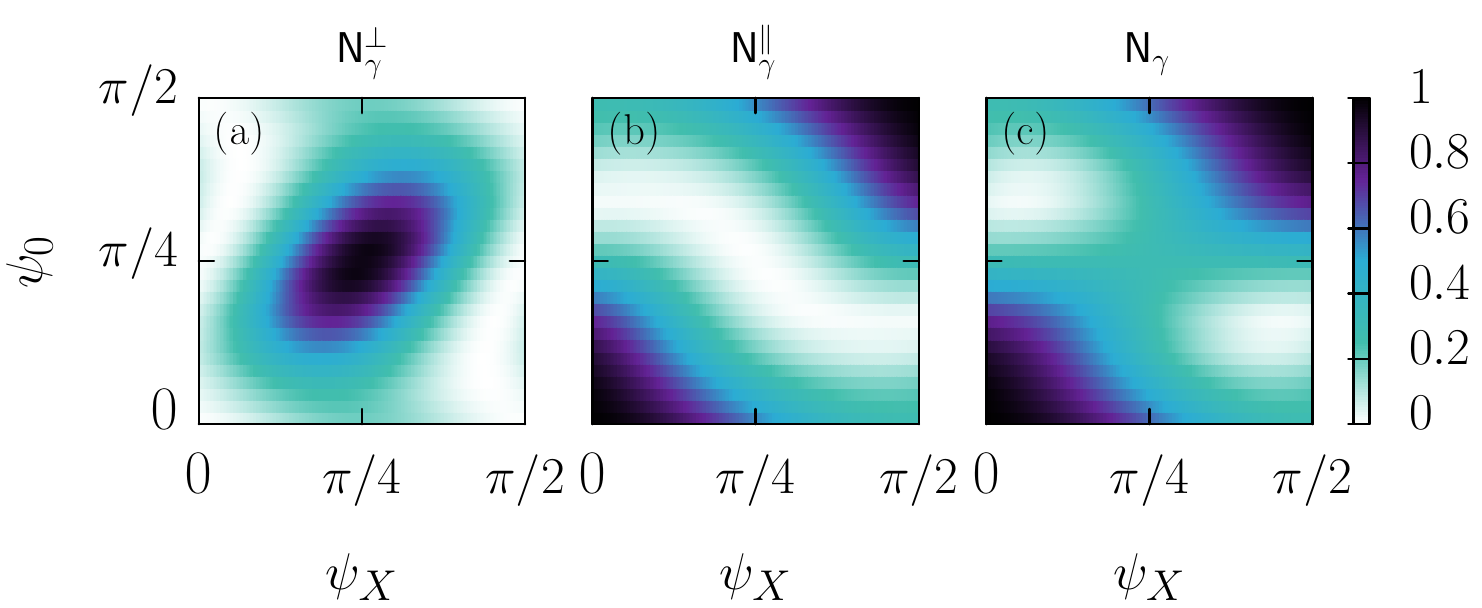}
    \caption{\label{fig:Prefactors} 
        Polarization dependence of NLO photon-photon scattering.
        The polarization plane of the x-ray beam is defined by angle the $\psi_{\mathsf{x}}$ [see~\eqnref{eqn:XFELparallel}] and the polarization plane of the optical lasers defined is by $\psi_{1} = \psi_{2} = \psi_{0}$ [see~\eqnref{eqn:Optical1pol}].
        Shown are the space-time independent prefactors of (a) $\tsf{N}_{\gamma}^{\parallel}$ and (b) $\tsf{N}_{\gamma}^{\perp}$, given by \eqnref{eqn:NumberApproxNLO}, and (c) $\tsf{N}_{\gamma} = \tsf{N}_{\gamma}^{\parallel} + \tsf{N}_{\gamma}^{\perp}$ (right-hand plot).
        Each plot has been normalized to a maximum of unity and \eqnref{eqn:Coefficients34} has been used.
    }
\end{figure}

\subsubsection*{Fundamental constants of six-photon scattering}

A highlight of the three-beam configuration is that $2n$-photon scattering is separated into different side peaks for different $n$ [see \eqnref{eqn:BraggPeak}].
If the NLO six-photon side peak is sufficiently separated from the LO side peak such that we can neglect the signal from the LO interaction, we find
\begin{align}
    \tsf{N}_{\gamma}^{\parallel}\approx
    &
    \big\{
        (3 \cc + \cd)
        \cos(2 \psi_{\tsf{x}})
        +
        2
        (3 \cc + \cd)
        \cos(2 \psi_{0})
        \nonumber \\
        &
        +
        3 (\cc-\cd) 
        \cos[2 (\psi_{\tsf{x}} - 2 \psi_{0})]
    \big\}^{2}
    \mathcal{I}_{H}
    \nonumber
    \,,
    \\\label{eqn:NumberApproxNLO}
    \tsf{N}_{\gamma}^{\perp}
    \approx
    &
    \big\{
        (3 \cc + \cd)
        \sin(2 \psi_{\tsf{x}}) 
        \nonumber \\ 
        & 
        +
        3 (\cc-\cd)
        \sin[2 (\psi_{\tsf{x}} - 2 \psi_{0})]
    \big\}^{2}
    \mathcal{I}_{H}
    \,,
\end{align}
and the space-time dependence can be factorised into the integral $\mathcal{I}_{H}$. Just as in the LO case [\eqnref{eqn:RatioLO}], we can form the ratio of polarized photons to arrive at a quantity that is relatively insensitive to shot-to-shot variations.

Alongside the six-photon fundamental low-energy constants~\cite{Ritus:1975pcc,Gies:2016yaa}, 
\begin{align}\label{eqn:Coefficients34}
    \cc
    =
    &
    8
    \bigg(
        1
        +
        \frac{2807}{480}
        \frac{\alpha}{\pi}
    \bigg)
    \,,
    \quad
    \cd
    =
    13
    \bigg(
        1
        +
        \frac{5033}{780}
        \frac{\alpha}{\pi}
    \bigg)
    \,,
\end{align}
the number of signal photons due to the NLO interaction depends explicitly on the polarization angles $\psi_{\tsf{x}}$ and $\psi_{0}$, rather than just their difference as was the case for the LO interaction \eqnref{eqn:NumberTotalApproxLO}.
The dependence on the polarization angles is illustrated in \figref{fig:Prefactors}.
The reason for this different dependence on the polarization is due to the configuration of colliding three beams. 
For the side peaks, the LO process of four-photon scattering has a different optical beam connected to the two pump vertices, whereas for the NLO process of six-photon scattering the four pump vertices involve the same optical beam being connected to two vertices. 
This has a restrictive effect on the NLO tensor structure, which makes it more sensitive to the beam polarizations.

From \figref{fig:Prefactors} it can be seen that the number of perpendicular polarized photons $\tsf{N}_{\gamma}^{\perp}$ is maximized when $\psi_{\tsf{x}} = \psi_{0} = \pi/4$, but, with this same choice, the number of parallel polarized photons $\tsf{N}_{\gamma}^{\parallel}$ goes to zero.
Likewise, the number of parallel polarized photons or total number of photons are maximized when $\psi_{\tsf{x}} = \psi_{0} = 0$ or $\pi/2$, where the number of perpendicular polarized photons $\tsf{N}_{\perp}$ falls to zero. 
Making use of the partial symmetry between $\psi_{\tsf{x}}$ and $\psi_{0}$, we simplify the analysis by choosing $\psi_{\tsf{x}} = \psi_{0} = \psi$. 
Using \eqnref{eqn:NumberApproxNLO}, we find
\begin{align}
    \tsf{N}_{\gamma}^{\parallel}
    \big|_{\psi_{\tsf{x}} = \psi_{0} = \psi}
    \approx
    &
    144
    \cc^{2}
    \cos^{2}(2 \psi)
    \mathcal{I}_{H}
    \,,
    \nonumber\\
    \tsf{N}_{\gamma}^{\perp}
    \big|_{\psi_{\tsf{x}} = \psi_{0} = \psi}
    \approx
    &
    16
    \cd^{2}
    \sin^{2}(2 \psi)
    \mathcal{I}_{H}
    \,,
    \label{eqn:NumberTotalApproxNLOSimple}
    \\\nonumber
    \tsf{N}_{\gamma}
    \big|_{\psi_{\tsf{x}} = \psi_{0} = \psi}
    \approx
    &
    8
    \big[
        9 \cc^{2}
        +
        \cd^{2}
        +
        (
        9 \cc^{2}
        -
        \cd^{2}
        )
        \cos(4\psi)
    \big]
    \mathcal{I}_{H}
    \,.
\end{align}
This also gives a simple ratio,
\begin{align}\label{eqn:RatioNLO}
    \frac{\tsf{N}_{\gamma}^{\perp}}{\tsf{N}_{\gamma}^{\parallel}}
    \Big|_{\psi_{\tsf{x}} = \psi_{0} = \psi}
    \approx
    \frac{\cd^{2}}{9 \cc^{2}}
    \tan^{2}(2 \psi)
    \,.
\end{align}

\section{Numerical results \label{sec:Results}}

In this section  we present numerical results for the number of scattered photons for two different sets of experimental parameters.
We consider a currently attainable case which could be sufficient to measure the LO process of four-photon scattering and a future parameters case which would allow access to the NLO process of six-photon scattering. 
Measurement of the NLO process of six-photon scattering is currently beyond experimental reach, but would be aided by having a more intense optical laser, higher brilliance XFEL beam, or collisions at a higher repetition rate.
The numerical results also act as a test of the analytically derived expressions relating the fundamental low-energy constants to numbers of polarized and unpolarized scattered photons.

\subsubsection*{Currently attainable parameters}

\begin{table*}
    \centering
    \begin{tabular}{ l c c c c c c c c c c c c }
        \hhline{=============}
        &
        \multicolumn{4}{ c }{Optical}
        &
        \multicolumn{4}{ c }{X-ray (Self-seeded)}
        &
        \multicolumn{4}{ c }{X-ray (SASE)}
        \\
        \hline
        Case
        &
        $\lambda_{0}$ ($\text{\textmu{}m}$) & $\tau_{0}$ ($\mathrm{fs}$) & $w_{0}$ ($\text{\textmu{}m}$) & $W$ ($\mathrm{J}$)  
        & 
        $\omega_{\tsf{x}}$ ($\mathrm{keV}$)  & $\tau_{\mathsf{x}}$ ($\mathrm{fs}$)  & $w_{\tsf{x}}$ ($\text{\textmu{}m}$)  & $\mathsf{N}_{\mathsf{x}}$ 
        & 
        $\omega_{\tsf{x}}$ ($\mathrm{keV}$)  & $\tau_{\mathsf{x}}$ ($\mathrm{fs}$)  & $w_{\tsf{x}}$ ($\text{\textmu{}m}$)  & $\mathsf{N}_{\mathsf{x}}$ 
        \\
        \hline
        EuXFEL~\cite{Ahmadiniaz:2024xob}
        &
        $0.8$ & $30$ & $\fnum \times 1.3$ & $4.9$ 
        & 
        $10$ & $25$ &  0.1-10 & $2 \times 10^{11}$
        & 
        $10$ & $25$ &  0.1-10 & $1 \times 10^{12}$
        \\
        SACLA~\cite{SACLA-yabuuchi2019experimental,SACLA-inoue2019generation}
        &
        $0.8$ & $25$ & $\fnum \times 1.3$ & $25$ 
        & 
        $10$ & $10$ &  0.1-10 & $1 \times 10^{11}$
        & 
        $10$ & $10$ &  0.1-10 & $4 \times 10^{11}$
        \\
        Future
        &
        $0.8$ & $30$ & $\fnum \times 1.3$ & ${\fnum}^{2} \times 4.2 \times 10^{3}$
        & 
        $10$ & $25$ &  0.1-10 & $2 \times 10^{11}$
        & 
        $10$ & $25$ &  0.1-10 & $1 \times 10^{12}$
        \\
        \hhline{=============}
    \end{tabular}
    \caption{
        \label{tab:Parameters}
        Optical and x-ray pulse parameters used in numerical results.
        The optical parameters are the wavelength $\lambda_{0}$, pulse duration $\tau_{0}$, beam waist $w_{0}$, focusing $f$-number $\fnum$, and total available pulse energy $W$.
        The x-ray parameters are the energy $\omega_{\mathsf{x}}$, pulse duration $\tau_{\mathsf{x}}$, beam waist $w_{\mathsf{x}}$, and photons per pulse $\mathsf{N}_{\mathsf{x}}$.
        Pulse duration is defined with respect to the FWHM value.
        The beam waist is defined as the value at which the intensity drops to $1/e^{2}$ of the peak.
    }
\end{table*}

For the currently attainable XFEL beam, technology is assumed that is capable of offering hard x-rays with energy $\omega_{\mathsf{x}} \sim O(10~\mathrm{keV})$ and numbers of photons per pulse $\mathsf{N}_{\mathsf{x}} \sim O(10^{11} - 10^{12})$.
The energy requirement ensures that the photon-photon cross section is enhanced relative to an all-optical configuration while the number of photons per pulse ensures that the number of signal photons, $\mathsf{N}_{\gamma}$, which scales as $\mathsf{N}_{\gamma} \propto \mathsf{N}_{\mathsf{x}}$, is large enough for a significant number of events to be observed. Also important is the availability of both SASE~\cite{Kondratenko.Part.Accel.1980,Bonifacio.Opt.Com.1984} and self-seeded~\cite{Feldhaus.Opt.Com.1997,Geloni.J.Mod.Opt.2011} operational modes.
For polarization resolved measurements, the self-seeded injection scheme is the most suitable (i.e. vacuum birefringence experiments), as typical crystal polarimeters have a narrow acceptance bandwidth $\Delta\omega_{\tsf{x}}$.
For polarization-insensitive measurements, one can instead use the SASE mode, which provides a larger number of photons per bunch at the expense of a larger bandwidth. For the currently attainable optical beams, we assume a peak power $P \sim O(\text{0.1-1\,PW})$ and short duration $\tau_{0} \sim O(10\,\mathrm{fs})$. The number of signal photons $\mathsf{N}_{\gamma}$ scales at LO with the optical beam intensity, $I$, as $\mathsf{N}_{\gamma} \propto I^{2}$.

Two examples of facilities meeting the above requirements are the European X-Ray Free-Electron Laser (EuXFEL)~\cite{Garcia.HPLSE.2021,Ahmadiniaz:2024xob} and the SPring-8 Angstrom Compact Free Electron Laser (SACLA)~\cite{SACLA-yabuuchi2019experimental,SACLA-inoue2019generation}.
The EuXFEL produces coherent pulses with photon energies in the range $8~\mathrm{keV} \lesssim \omega_{\tsf{x}} \lesssim 30~\mathrm{keV}$.
These can be combined with the Ti:sapphire optical ReLaX laser system installed at the Helmholtz International Beamline for Extreme Fields (HIBEF)~\cite{Garcia.HPLSE.2021}, offering a total pulse energy of $W = 4.9~\mathrm{J}$ at a $\tau_{0} = 30~\mathrm{fs}$ duration, corresponding to a peak power $P \approx 200~\mathrm{TW}$.
The waist is given by $w_{0} = \fnum \times 1.3~\text{\textmu{}m}$, where $\fnum$ is the f-number of the focusing optics.
Assuming that the ReLaX beamline is passed through a 50:50 beam splitter to produce two optical pulses with energy $W/2 = 2.45~\mathrm{J}$, for $\fnum = 1$ each pulse can reach an intensity of $I_{0} \approx 5.7 \times 10^{21}~\mathrm{Wcm^{-2}}$.
The EuXFEL parameters in \tabref{tab:Parameters} are taken to coincide with those of the proposed BIREF@HIBEF experiment~\cite{Ahmadiniaz:2024xob}.
SACLA offers a similar XFEL capability, providing coherent pulses with photon energies $4~\mathrm{keV} \lesssim \omega_{\tsf{x}} \lesssim 15~\mathrm{keV}$.
At SACLA there is also a Ti:Sa optical laser system capable of delivering two $\tau_{0} = 25~\mathrm{fs}$ duration laser pulses with combined total energy $W = 25~\mathrm{J}$, corresponding to a total peak power $P \approx 1~\mathrm{PW}$~\cite{SACLA-yabuuchi2019experimental}.
Assuming the capability to focus each pulse to a beam waist $w_{0} = \fnum \times 1.3~\text{\textmu{}m}$, as for the ReLaX system above, this corresponds to a peak intensity of $I_{0} \approx 3.5 \times 10^{22}~\mathrm{Wcm^{-2}}$ per pulse.
In the following sections we will evaluate the feasibility of performing photon-photon scattering discovery experiments at both EuXFEL and SACLA.

\subsubsection*{Future Parameters}

Going beyond the constraints of currently available technology we also consider what sort of signals one may expect at next generation facilities.
Our goal is to explore the feasibility of using such a set up to access both the LO and NLO photon-photon scattering contributions.
The number of signal photons from the NLO contribution will be proportional to $\tsf{N}_{\gamma} \propto E_{\mathsf{x}}^{2} E_{0}^{8}$, such that there is a strong dependence on the field strength of the optical laser pulses.
For this reason, we focus on the interaction of a next generation multi-kJ-class high-power laser system with x-ray pulses.
Many x-ray free-electron laser facilities which provide high-brightness hard x-ray pulses operate with a similar capability to the EuXFEL and so we will use these parameters for the x-ray beam as a demonstrative example.
For the optical laser, we consider pulses of 30~fs duration generated from a Ti:sapphire laser system with total energy $W$ such that for fixed $f^{{\#}}$ the intensity of each pulse is $I_{0} = E_{\mathsf{cr}}^{2} \mathcal{E}_{0}^{2} = 10^{{25}}~\mathrm{Wcm^{-2}}$ [using \eqnref{eqn:FieldStrengths}].
The parameters of such a laser system are outlined in \tabref{tab:Parameters}.

\subsection{Background estimation \label{sec:Background}}

The number of signal photons per optimal collision of the three beams can reach $\mathsf{N}_{\gamma} \sim O(10^{-5}\text{-}10^{-1})$ per collision for the currently attainable parameters in \tabref{tab:Parameters}. 
The number of x-ray photons per shot is approximately $O(10^{11}\text{-}10^{12})$. 
Although the signal is separated from the background in spatial co-ordinate and polarization, a good understanding of the background is essential in any experiment. 
Considering the background also allows one to determine suitable values for the remaining free x-ray and optical beam parameters, namely the x-ray photon energy $\omega_{\mathsf{x}}$ and the x-ray and optical beam waists $w_{\mathsf{x}}$ and $w_{0}$, respectively.

The majority of the XFEL  photons will be filtered out by placing the detector at the position of the Bragg side peaks and ensuring those peaks are a suitable distance from the central peak. 
Consider a detector a distance $L \sim O(\mathrm{m})$ down the x-ray propagation axis with a circular exclusion region of radius $R$, corresponding to an angular cut $\theta_{\ast} = \tan^{-1}(R/L)$.
Assuming that the detector can measure all signal photons with $\theta_{\gamma} > \theta_{\ast}$, the total number of signal photons will be
\begin{align}\label{eqn:Nstar}
    \mathsf{N}_{\gamma,\ast}
    =
    \int_{\theta_{\ast}}^{\infty} 
    \frac{\ud \mathsf{N}_{\gamma}}{\ud \theta_{\gamma}}
    \ud \theta_{\gamma}
    \,.
\end{align}
Treating the measurement of the number of scattered photons as Poissonian with a background count per shot of $\mathsf{N}_{\gamma}^{\mathsf{bg}}$, following \cite{Cowan:2010js}, the number of optimal shots required to reach a statistical significance of $\mathsf{n}\sigma$ using the three-beam scenario is
\begin{align}\label{eqn:Shots}
    \mathsf{N}_{\mathsf{shots}}^{\mathsf{n}\sigma}
    \gtrsim
    &
    \frac{\mathsf{n}^{2}}{2}
    \Big[
        (\mathsf{N}_{\gamma, \ast} + \mathsf{N}_{\gamma}^{\mathsf{bg}}) 
        \ln \Big(
            1 + \frac{\mathsf{N}_{\gamma, \ast}}{\mathsf{N}_{\gamma}^{\mathsf{bg}}}
        \Big) 
        - 
        \mathsf{N}_{\gamma, \ast}
    \Big]^{-1}
    \,.
\end{align}

We identify two potential sources of background for all measurements: the divergence of the x-ray beam, $\mathsf{N}_{\mathsf{x}}^{\mathsf{bg}}$, and Compton scattering from impurities in the vacuum chamber, $\mathsf{N}_{\mathsf{C}}^{\mathsf{bg}}$.
Another source of background, effecting only polarization-sensitive measurements, is due to the polarization purity of the probe x-ray pulse, $\mathsf{N}_{\mathcal{P}}^{\mathsf{bg}}$.
The total background is then
\begin{align}\label{eqn:Background}
    \mathsf{N}_{\gamma}^{\mathsf{bg}}
    =
    \mathsf{N}_{\mathsf{x}}^{\mathsf{bg}}
    +
    \mathsf{N}_{\mathsf{C}}^{\mathsf{bg}}
    +
    \mathsf{N}_{\mathcal{P}}^{\mathsf{bg}}
    \,.
\end{align}

The optical elements used to focus the x-ray beam will also impact the number of signal and background photons per shot.
Assuming the use of two standard beryllium lenses, one to focus the probe beam and another to focus the signal, each with a $50\%$ transmission, this leads to a total reduction by a factor $0.25$ in the number of signal and background photons.

\paragraph*{X-ray beam divergence: }
The biggest potential source of background comes from the divergence of the probe x-ray beam.
Since $\mathsf{N}_{\mathsf{x}} \gg \mathsf{N}_{\gamma,\ast}$, measurements will be background dominated if even a fraction of a percent of probe photons diverge beyond the detector exclusion region.
The number of XFEL photons beyond the angular cut $\theta_{\ast}$ can be estimated by integrating $E_{\tsf{x}}^2$, given by \eqnref{eqn:Gaussian}, on the detector over time and azimuthal angle
\begin{align}\label{eqn:NxBG}
\mathsf{N}_{\mathsf{x}}^{\mathsf{bg}}
    \approx
    \mathsf{N}_{\mathsf{x}}\mbox{e}^{ - \sigma_{\tsf{x}}^{-2}         \tan^{2} \theta_{\ast}};        \quad
    \sigma_{\tsf{x}}^{2} = \frac{w_{\tsf{x}}^2}{2L^{2}}+\frac{1}{2f_{\tsf{x}}^{2}\pi^2}    
\end{align}
where $f_{\tsf{x}}=w_{\tsf{x}}/\lambda_{\tsf{x}}$ is the focusing $f_{\mathsf{x}}$-number and $\lambda_{\mathsf{x}} = 2\pi/\omega_{\mathsf{x}}$ is the wavelength of the x-rays. 
If one demands $\mathsf{N}_{\mathsf{x}}^{\mathsf{bg}}\ll\tsf{N}_{\gamma}$, then:
\begin{align}\label{eqn:AngleCondition1}
    \tan \theta_{\ast} \gg \sigma_{\ast}\,,
    \quad 
    \sigma_{\ast}
    =
    \sigma_{\tsf{x}} \left(\ln \frac{\mathsf{N}_{\tsf{x}}}{\mathsf{N}_{\gamma}}\right)^{1/2}
    \,.
\end{align}
For the XFEL plus optical setup considered here, $\tan \theta_{\ast} \approx \theta_{\ast}$ and $\sigma_{\tsf{x}} \approx 1/f_{\tsf{x}}\pi\sqrt{2}$. 
If we choose $\theta_{\ast} > C \sigma_{\tsf{x}}$ for some $C$ to fulfill \eqnref{eqn:AngleCondition1}, then a condition on the angular detector cut is:
\begin{align}\label{eqn:AngleCondition2}
    \theta_{\ast} > \frac{C}{f_{\tsf{x}}\pi\sqrt{2}}
    \,.
\end{align}
We expect that when the signal is largest, the center of the Bragg side peaks will be on the detector, $\bragg > \theta_{\ast}$, which also implies $\bragg > C \sigma_{\mathsf{x}}$.
Using \eqnref{eqn:BraggPeak} this gives a condition:
\begin{align}\label{eqn:WaistCondition}
    w_{\tsf{x}} > \frac{C \lambda_{0}}{2 \sqrt{2} \pi (n - 1) \sin\Theta}
    \,.
\end{align}

These conditions imply a hierarchy for the associated angles, $\theta_{\mathsf{x}} < \theta_{\ast} < \theta_{\gamma}$.
Considering the $C = 10$ standard deviation of an x-ray pulse with $\omega_{\mathsf{x}} = 10~\mathrm{keV} \implies \lambda_{\mathsf{x}} = 0.12~\mathrm{nm}$, the angular cut must satisfy $\theta_{\ast} (\text{\textmu{}rad}) > 270 w_{\mathsf{x}}^{-1}(\text{\textmu{}m})$.
Similarly, considering the optical pulse with $\lambda_{0} = 0.8\,\text{\textmu{}m}$ and a collision angle of $\Theta = 30^{\circ}$, the condition in \eqnref{eqn:WaistCondition} sets a lower bound of $w_{\mathsf{x}} > 1.8\,\text{\textmu{}m}$ for the LO contribution with $n = 2$.
Thus, a choice of $w_{\mathsf{x}} = 4\,\text{\textmu{}m}$ and $\theta_{\ast} = 200\,\text{\textmu{}rad}$ satisfies the x-ray divergence and Bragg side peak conditions, and a detector placed $L = 5\,\mathrm{m}$ down the x-ray beamline will measure, from \eqnref{eqn:NxBG}, $\mathsf{N}_{\mathsf{x}}^{\mathsf{bg}} \approx e^{-870}  \approx 0$ background photons from the x-ray beam divergence outside the exclusion region, such that we can neglect this contribution to the background.

\paragraph*{Compton scattering: }
Another source of background will be from Compton scattering of x-ray photons and electrons in the interaction vacuum chamber, as in practice a perfect vacuum is not feasible. 
The interaction geometry of the three-beam setup makes a precise calculation of the background due to Compton scattering a nontrivial task \cite{Ahmadiniaz:2022mcy}.
However, one can estimate an upper bound on this background in the following way.
Given an x-ray pulse with $\mathsf{N}_{\mathsf{x}}$ photons in a duration $\tau$, the number of scattered photons due to Compton scattering can be estimated as $\mathsf{N}_{\mathsf{C}}^{\mathsf{bg}} = \sigma_{\mathsf{T}} n_{e} \mathsf{N}_{\mathsf{x}} \tau$, where $\sigma_{\mathsf{T}} = 6.65 \times 10^{-29}~\mathrm{m}^{2}$ is the Thomson scattering cross section and $n_{e}$ is the density of electrons.
We assume a vacuum of $10^{-9}~\mathrm{torr}$, corresponding to $n_{e} \approx 5 \times 10^{9}~\mathrm{cm}^{-3}$, which gives $\mathsf{N}_{\mathsf{C}}^{\mathsf{bg}} \approx \mathsf{N}_{\mathsf{x}} \times 2.5 \times 10^{-18}$.
This estimate does not include considerations about the number of these photons which will reach the detectors or the effect of ponderomotive blow out on the electron population, both of which should act to reduce the contribution to the background from Compton scattering (see, e.g.,~\cite{Lundstrom:2005za} for a similar discussion in the context of colliding three optical pulses).
Thus, $\mathsf{N}_{\mathsf{C}}^{\mathsf{bg}}$ can be viewed as an upper bound.

\paragraph*{X-ray polarization purity: }
For polarization-sensitive measurements we must also consider the effect of the polarization purity of the probe x-ray beam, $\mathcal{P} = \mathsf{N}_{\mathsf{x}}^{\perp}/\mathsf{N}_{\mathsf{x}}$, where $\mathsf{N}_{\mathsf{x}}^{\perp}$ is the number of photons in the probe pulse which are polarized in the orthogonal polarization mode.
An orthogonally polarized photon in the initial beam scattering without polarization flip will contribute a background $\mathsf{N}_{\mathcal{P}}^{\mathsf{bg} \perp} \approx \mathcal{P} \mathsf{N}_{\gamma,\ast}^{\parallel}$ to the $\mathsf{N}_{\gamma,\ast}^{\perp}$ measurement, while those photons scattering with polarization flip will contribute a background  $\mathsf{N}_{\mathcal{P}}^{\mathsf{bg} \parallel} \approx \mathcal{P} \mathsf{N}_{\gamma,\ast}^{\perp}$ to the $\mathsf{N}_{\gamma,\ast}^{\parallel}$ measurement. 
While this background could be reduced by using additional polarisers to improve the polarization purity of the x-ray pulses, these also introduce detrimental effects such as transmission losses and pulse lengthening~\cite{Karbstein:2021ldz}. 
We will find below that the EuXFEL polarization purity of $\mathcal{P} \approx 10^{-6}$ without additional polarisers leads to backgrounds that are sufficiently small \cite{Geloni.Opt.Comms.2015}.
For polarization-insensitive measurements, where only the photon count on the detector is important, the background $\mathsf{N}_{\mathcal{P}}^{\mathsf{bg}}$ is a component of the signal itself, and so can be neglected.

\subsection{Polarization-sensitive measurements \label{sec:CurrentpolarizationSensitive}}

\subsubsection{Currently available technology: four-photon scattering\label{sec:Current}}

\begin{figure}
    \includegraphics[width=0.45\textwidth,trim={0.0cm 0.0cm 0.0cm 0.0cm},clip=true]{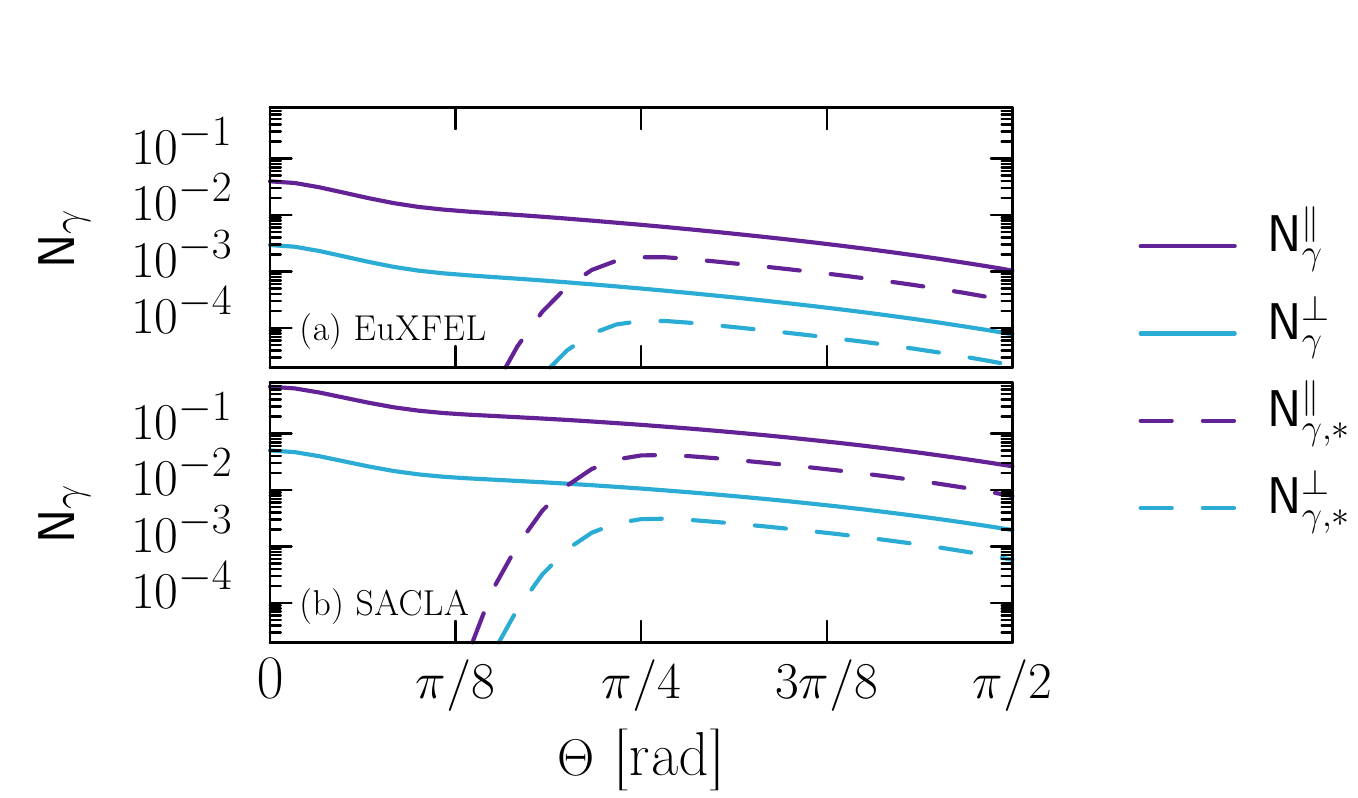}
    \caption{\label{fig:ModesNumberAngleCurrent} 
        Number of signal photons vs collision angle $\Theta$ using $\omega_{\tsf{x}} = 10~\mathrm{keV}$ self-seeded (a) EuXFEL parameters and (b) SACLA parameters.
        Optical pulses have $\fnum = 1$ focusing, x-ray pulses are focused to $w_{\tsf{x}} = 4~\text{\textmu{}m}$, and the relative polarization angle is $\Delta\psi = \pi/4$. 
        Plotted are the number of photon $\tsf{N}_{\gamma}^{\parallel}$ scattered into the parallel state  (purple solid line), the number $\tsf{N}_{\gamma}^{\perp}$ scattered into the perpendicular state (blue solid line), and the equivalent photon counts if an angular cut is applied which only accepts photons with emission angles $\theta_{x} > 200~\text{\textmu{}rad}$ (corresponding dashed lines).
    }
\end{figure}

Our goal now is to estimate the number of signal photons which could be obtained in a polarization-sensitive measurement using technology currently available at EuXFEL and SACLA, using \tabref{tab:Parameters}.
We consider the scattering of $\omega_{\mathsf{x}} = 10~\mathrm{keV}$ self-seeded photons with optical pulses at $\fnum = 1$ ($w_{0} = 1.3~\text{\textmu{}m}$) focusing.
The remaining free parameters are $C = 10$, $w_{\mathsf{x}} = 4~\text{\textmu{}m}$ and $\theta_{\ast} = 200~\text{\textmu{}rad}$, satisfying the constraints \eqnref{eqn:AngleCondition1}--\eqref{eqn:WaistCondition}.
The angle between the x-ray and optical polarization planes is chosen as $\Delta\psi=\pi/4$, maximizing the number of x-ray photons scattered into the flipped polarization state [c.f.~\eqnref{eqn:NumberTotalApproxLO}].
The number of signal photons per optimal shot  neglecting losses due to x-ray optics is shown in \figref{fig:ModesNumberAngleCurrent} for both EuXFEL [\figref{fig:ModesNumberAngleCurrent}(a)] and SACLA [\figref{fig:ModesNumberAngleCurrent}(b)] parameters with $\fnum = 1$ focusing of the optical pulses.
Choosing an angular cut $\theta_{\ast} = 200~\text{\textmu{}rad}$, the dashed lines in \figref{fig:ModesNumberAngleCurrent} show the number of parallel ($\tsf{N}_{\gamma,\ast}^{\parallel}$) and perpendicular ($\tsf{N}_{\gamma,\ast}^{\perp}$) polarized photons with scattering angles $\theta_{\gamma} > \theta_{\ast}$, peaking at a collision angle $\Theta \approx 45^{\circ}$ with $\tsf{N}_{\gamma,\ast}^{\parallel} \approx 1.8 \times 10^{-3}$ and $\tsf{N}_{\gamma,\ast}^{\perp} \approx 1.3 \times 10^{-4}$ for EuXFEL parameters and at $\Theta \approx 48^{\circ}$ with $\tsf{N}_{\gamma,\ast}^{\parallel} \approx 4.2 \times 10^{-2}$ and $\tsf{N}_{\gamma,\ast}^{\perp} \approx 3.1 \times 10^{-3}$ for SACLA parameters.

The number of background photons of any polarization due to Compton scattering is determined as $\mathsf{N}_{\mathsf{C}}^{\mathsf{bg}} \approx 5 \times 10^{-7}$ for the EuXFEL parameters and for $\mathsf{N}_{\mathsf{C}}^{\mathsf{bg}} \approx 2.5 \times 10^{-7}$ SACLA parameters. 
The number of background photons in a specific polarization can be acquired by considering x-ray pulse polarization purity, $\mathcal{P} = \mathsf{N}_{\mathsf{x}}^{\perp} / \mathsf{N}_{\mathsf{x}} \approx 10^{-6}$~\cite{Geloni.Opt.Comms.2015}, giving $\mathsf{N}_{\mathsf{C}}^{\mathsf{bg}\parallel} = (1 - \mathcal{P}) \mathsf{N}_{\mathsf{C}}^{\mathsf{bg}}$ for the Compton parallel polarization background, and $\mathsf{N}_{\mathsf{C}}^{\mathsf{bg}\perp} = \mathcal{P} \mathsf{N}_{\mathsf{C}}^{\mathsf{bg}}$ for the Compton perpendicular polarization background.
This corresponds to $(\mathsf{N}_{\mathsf{C}}^{\mathsf{bg}\parallel},\mathsf{N}_{\mathsf{C}}^{\mathsf{bg}\perp}) = (5 \times 10^{-7},5 \times 10^{-13})$ for the EuXFEL and $(\mathsf{N}_{\mathsf{C}}^{\mathsf{bg}\parallel},\mathsf{N}_{\mathsf{C}}^{\mathsf{bg}\perp}) = (2.5 \times 10^{-7},2.5 \times 10^{-13})$ for SACLA.
For polarization-sensitive measurements we also have to estimate the direct contribution to the background from the polarization purity.
For the total number of x-ray photons per pulse of $\mathsf{N}_{\mathsf{x}}$, there are  $\mathsf{N}_{\mathsf{x}}^{\perp} = \mathcal{P} \mathsf{N}_{\mathsf{x}}$ probe photons in the orthogonal polarization mode.
The contribution to the background from these photons depends on the pulse parameters and configuration.
Considering the collision at $\Theta = 45^{\circ}$ which approximately maximizes the signals for photons with $\theta_{\gamma} > \theta_{\ast}$ for EuXFEL and SACLA parameters, we find the polarization purity contributes $\mathsf{N}_{\mathcal{P}}^{\mathsf{bg}\parallel} \approx 1.3 \times 10^{-10}$ and
$\mathsf{N}_{\mathcal{P}}^{\mathsf{bg}\perp} \approx 1.8 \times 10^{-9}$  background photons to, respectively, the parallel and perpendicular EuXFEL measurements, and  $\mathsf{N}_{\mathcal{P}}^{\mathsf{bg}\parallel} \approx 3.0 \times 10^{-9}$ and
$\mathsf{N}_{\mathcal{P}}^{\mathsf{bg}\perp} \approx 4.1 \times 10^{-8}$  background photons to the SACLA measurements.
Thus, the main contribution to the parallel polarization background is from Compton scattering, while the polarization purity of the probe is the dominant background factor of the perpendicular measurement.
Including the effect of focusing optics by reducing the signal and background contributions by a factor $0.25$, the detected photon counts are given in \tabref{tab:BirefringentNumbers}.
Using these in \eqnref{eqn:Shots}, we can estimate the number of optimal shots which would be required to achieve a statistical significance of $5\sigma$ in a polarization-sensitive measurement.
For EuXFEL parameters we find $\mathsf{N}_{\mathsf{shots}}^{5\sigma,\parallel} \gtrsim 3.9 \times 10^{3}$ and $\mathsf{N}_{\mathsf{shots}}^{5\sigma,\perp} \gtrsim 3.7 \times 10^{4}$, while for SACLA parameters we find $\mathsf{N}_{\mathsf{shots}}^{5\sigma,\parallel} \gtrsim 1.2 \times 10^{2}$ and $\mathsf{N}_{\mathsf{shots}}^{5\sigma,\perp} \gtrsim 1.6 \times 10^{3}$.
Since the parallel and perpendicular photons can be measured simultaneously in a single shot, the total number of required shots to obtain $5\sigma$ significance in both observables will be determined by $\mathsf{N}_{\mathsf{shots}}^{5\sigma} \gtrsim \mathrm{max}(\mathsf{N}_{\mathsf{shots}}^{5\sigma,\parallel},\mathsf{N}_{\mathsf{shots}}^{5\sigma,\perp})$, which is also shown in \tabref{tab:BirefringentNumbers}.

\begin{figure}[t!!]
    \includegraphics[width=0.45\textwidth,trim={0.25cm 0.0cm 0.0cm 0.5cm},clip=true]{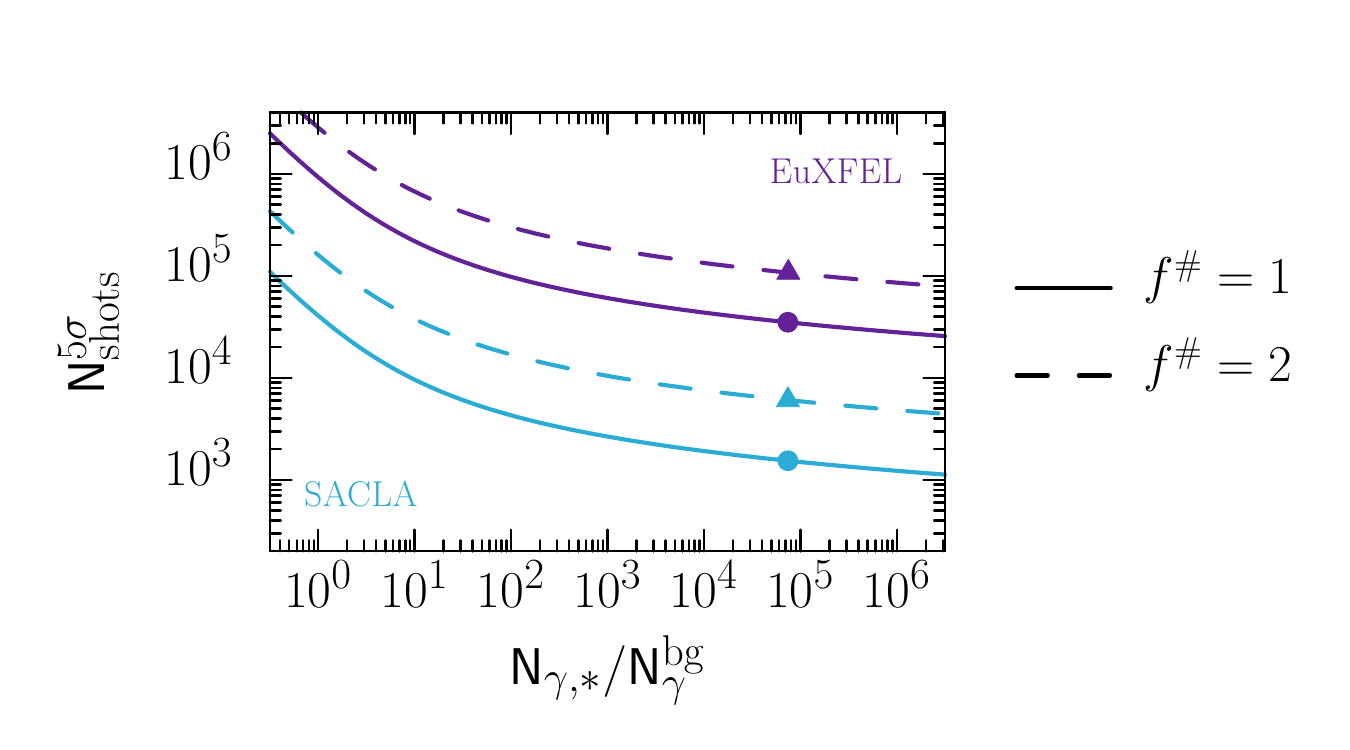}
    \caption{\label{fig:ModesShots} 
        Estimated lower bound on the number of shots  required for $5\sigma$ statistical significance, $\mathsf{N}_{\mathsf{shots}}^{5\sigma} = \mathrm{max}(\mathsf{N}_{\mathsf{shots}}^{5\sigma,\parallel},\mathsf{N}_{\mathsf{shots}}^{5\sigma,\perp})$, as a function of the signal-to-background ratio $\mathsf{N}_{\gamma,\ast}/\mathsf{N}_{\gamma}^{\mathsf{bg}}$.
        The x-ray and optical pulse parameters are as in \figref{fig:ModesNumberAngleCurrent} with $\Theta = 45^{\circ}$.
        Estimations are of the number of signal photons held fixed and are given by \tabref{tab:BirefringentNumbers}.
        The purple solid line shows $\fnum = 1$ (EuXFEL);
        purple dashed line, $\fnum = 2$ (EuXFEL);
        blue solid line, $\fnum = 1$ (SACLA);
        and blue dashed line, $\fnum = 2$ (SACLA).
    The data points correspond to the background estimations in \tabref{tab:BirefringentNumbers}.
    }
\end{figure}

\begin{table*}
    \centering
    {\setlength{\tabcolsep}{1em}
    \begin{tabular}{l c c c c c c c }
        \hhline{========}
        Case
        &
        $\fnum$
        &
        $\mathsf{N}_{\gamma,\ast}^{\parallel}$ 
        & 
        $\mathsf{N}_{\gamma}^{\mathsf{bg}\parallel}$ 
        & 
        $\mathsf{N}_{\gamma,\ast}^{\perp}$ 
        & 
        $\mathsf{N}_{\gamma}^{\mathsf{bg}\perp}$ 
        &
        $\mathsf{N}_{\mathsf{shots}}^{3\sigma}$ 
        &
        $\mathsf{N}_{\mathsf{shots}}^{5\sigma}$ 
        \\
        \hline
        EuXFEL
        &
        1
        &
        $4.5 \times 10^{-4}$
        &
        $1.3 \times 10^{-7}$
        &
        $3.3 \times 10^{-5}$
        &
        $4.5 \times 10^{-10}$
        &
        $1.3 \times 10^{4}$
        &
        $3.7 \times 10^{4}$
        \\
        EuXFEL
        &
        2
        &
        $1.5 \times 10^{-4}$
        &
        $1.3 \times 10^{-7}$
        &
        $1.1 \times 10^{-5}$
        &
        $1.5 \times 10^{-10}$
        &
        $4.0 \times 10^{5}$
        &
        $1.1 \times 10^{5}$
        \\
        SACLA
        &
        1
        &
        $1.0 \times 10^{-2}$
        &
        $6.3 \times 10^{-8}$
        &
        $7.6 \times 10^{-4}$
        &
        $1.0 \times 10^{-8}$
        &
        $5.8 \times 10^{2}$
        &
        $1.6 \times 10^{3}$
        \\
        SACLA
        &
        2
        &
        $2.6 \times 10^{-3}$
        &
        $6.3 \times 10^{-8}$
        &
        $1.9 \times 10^{-4}$
        &
        $2.6 \times 10^{-9}$
        &
        $2.3 \times 10^{3}$
        &
        $6.3 \times 10^{3}$
        \\
        \hhline{========}
    \end{tabular}
    }
    \caption{\label{tab:BirefringentNumbers} 
        Estimated number of parallel, $\mathsf{N}_{\gamma,\ast}^{\parallel}$, and perpendicular, $\mathsf{N}_{\gamma,\ast}^{\perp}$, polarized signal photons in a polarization-sensitive measurement with associated backgrounds, $\mathsf{N}_{\gamma}^{\mathsf{bg}\parallel}$ and  $\mathsf{N}_{\gamma}^{\mathsf{bg}\perp}$.
        Estimates account for the use of two standard beryllium lenses with 50\% transmission to focus the probe and scattered beams.
        Also shown is the estimated lower bound on the number of shots  required for $3\sigma$ and $5\sigma$ statistical significance, $\mathsf{N}_{\mathsf{shots}}^{\mathsf{n}\sigma} = \mathrm{max}(\mathsf{N}_{\mathsf{shots}}^{\mathsf{n}\sigma,\parallel},\mathsf{N}_{\mathsf{shots}}^{\mathsf{n}\sigma,\perp})$, calculated assuming an optimal collision.
        Therefore it is an approximate lower bound on the number of shots required as effects such as beam jitter can reduce the number of signal photons.
    }
\end{table*}

While we have provided estimates and chosen suitable parameters to minimize the background contributions for idealized shots, we have not taken into account experimental fluctuations.
The estimates of the number of shots required for a given $\mathsf{n}\sigma$ will be sensitive to the actual background which is measured in each shot.
When more factors are taken into account, the level of the background in experiment may take different values than considered here. Therefore, it is useful to understand how the projected number of required shots changes with the background, for the estimated number of signal photons in \tabref{tab:BirefringentNumbers}.
In \figref{fig:ModesShots} the lower bound of the number of shots required for a statistical significance of $5\sigma$ is calculated according to $\mathsf{N}_{\mathsf{shots}}^{5\sigma} = \mathrm{max}(\mathsf{N}_{\mathsf{shots}}^{5\sigma,\parallel},\mathsf{N}_{\mathsf{shots}}^{5\sigma,\perp})$, using \eqnref{eqn:Shots}, as a function of the signal-to-background ratio $\mathsf{N}_{\gamma,\ast}/\mathsf{N}_{\gamma}^{\mathsf{bg}}$.
Since the number of signal photons in each case is fixed to the estimates in \tabref{tab:BirefringentNumbers}, an increase in $\mathsf{N}_{\gamma,\ast}/\mathsf{N}_{\gamma}^{\mathsf{bg}}$ corresponds to a decrease in the background.
The data points correspond to the background estimations in \tabref{tab:BirefringentNumbers}. 

\begin{figure}
    \includegraphics[width=0.45\textwidth,trim={0.0cm 0.0cm 0.0cm 0.0cm},clip=true]{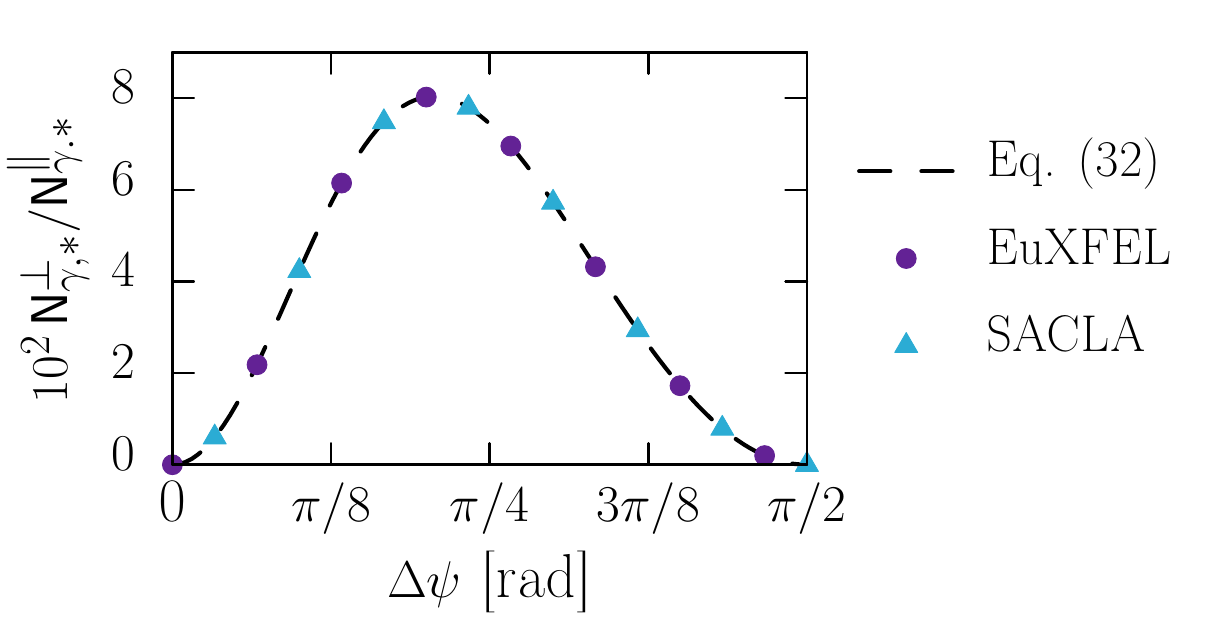}
    \caption{\label{fig:ModesRatioPolarisationCurrent} 
        Ratio of the number of signal photons in each polarization mode $\tsf{N}_{\gamma,\ast}^{\perp}/\tsf{N}_{\gamma,\ast}^{\parallel}$ vs relative polarization $\Delta\psi$ for a fixed collision angle $\Theta = 45^{\circ}$, $\fnum = 1$ optical focusing, and $w_{\mathsf{x}} = 4~\text{\textmu{}m}$.
        The black dashed line shows the analytical result from \eqnref{eqn:RatioLO} with $\ca$ and $\cb$ given by \eqnref{eqn:Coefficients12},
        purple circles show the numerically evaluated ratio using EuXFEL parameters,
        and blue triangles show the numerically evaluated ratio using SACLA parameters.
    }
\end{figure}

In \secref{sec:Constants} it was shown that the ratio of polarized scattered photons, $\tsf{N}_{\gamma}^{\perp}/\tsf{N}_{\gamma}^{\parallel}$, is independent of the space-time overlap of the beams. Determining the ratio allows the fundamental QED low-energy constants to be inferred. This conclusion is supported by the numerical result in \figref{fig:ModesRatioPolarisationCurrent}, which calculates this ratio numerically and finds the same result as the analytical prediction \eqnref{eqn:Coefficients12}.
This also confirms the dependency of the ratio of polarized photons $\tsf{N}_{\gamma}^{\perp}/\tsf{N}_{\gamma}^{\parallel}$ on
the angle $\Delta \psi$ between the x-ray and optical polarization planes from \eqnref{eqn:RatioLO}.

\begin{figure}
    \includegraphics[width=0.45\textwidth,trim={0.0cm 0.0cm 0.0cm 0.0cm},clip=true]{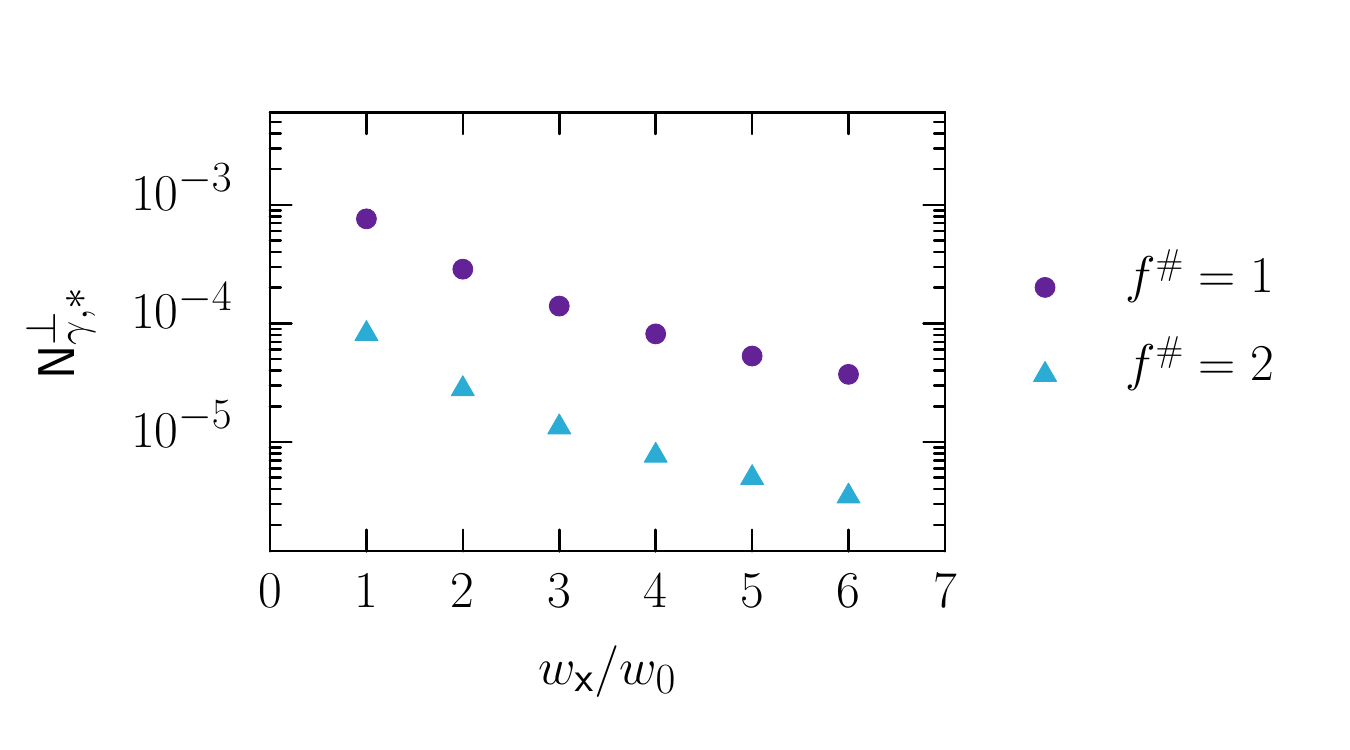}
    \caption{\label{fig:WaistCurrent} 
        Dependence of the number of perpendicular polarized signal photons $\tsf{N}_{\gamma,\ast}^{\perp}$ at collision angle $\Theta = 45^{\circ}$ on the ratio of the x-ray and optical beam waists $w_{\tsf{x}}/w_{0}$ using EuXFEL parameters with  $\fnum = 1$ (purple circles) and $\fnum = 2$ focusing (blue triangles).
    }
\end{figure}

In experiment it may be advantageous to increase the x-ray and optical beam-overlap (for example, to minimize the effect of beam jitter), which can be achieved by defocusing the optical laser. 
The number of polarization-flipped photons $\tsf{N}_{\gamma}^{\perp} \propto I_{\tsf{x}} I_{0}^{2}$, where $I$ is the intensity of the corresponding field changing the optical f-number from $\fnum = 1$ to $\fnum = 2$, naively results in the intensity change $I_{0} \to I_{0} / 4$ and the reduction in the number of signal photons by over a magnitude $\tsf{N}_{\gamma}^{\perp} \to \tsf{N}_{\gamma}^{\perp} / 16$.
However, if $w_{\mathsf{x}}$ is kept fixed and $w_{\mathsf{x}} > w_{0}$, the reduction due to the increase in $w_{0}$ will be partially compensated by more probe photons being within the foci of the optical pulses.
Figure \ref{fig:WaistCurrent} shows how the birefringence signal depends on the relative overlap of the x-ray and optical beams, which can be quantified by the ratio $w_{\tsf{x}}/w_{0}$, using EuXFEL parameters.
Overall, we see for fixed $w_{\tsf{x}}/w_{0}$ that doubling $\fnum$ leads to around an order of magnitude reduction in $\tsf{N}_{\gamma}^{\perp}$.
For fixed $\fnum$, going from $w_{\tsf{x}}/w_{0} = 1$ to $w_{\tsf{x}}/w_{0} = 6$ results in around an order of magnitude decrease in the number of scattered photons.

\subsubsection{Next generation technology: six-photon scattering\label{sec:FuturepolarizationSensitive}}

With each power of the electric-field strength in $\mathcal{L}$ in \eqnref{eqn:LagrangianExpand} there is also an additional suppression by the critical field strength $\Ecr$.
Therefore, the measurable photons mostly originate from the LO process; when attempting to isolate the NLO process, the LO process becomes a source of background.
One tool that we can use to aid with separating the contributions is the angular cut on the detector. 
The Bragg peaks corresponding to the NLO interaction are centered around $\theta_{x} \approx \theta_{\mathsf{B},6} = 4 \omega_{0}/\omega_{\mathsf{x}} \sin\Theta$, which is twice as far from the origin as the LO Bragg peak [see \eqnref{eqn:BraggPeak}].
Thus, we expect that a larger angular cut (which will turn out to be around twice the cut used in the preceding section) can be taken to isolate the signal from the NLO interaction on the detector.
This has an additional effect of allowing us to choose a smaller beam waist for the x-ray pulses $w_{\mathsf{x}}$ while still ensuring that the background due to its divergence is negligible and the conditions \eqnref{eqn:NxBG}--\eqref{eqn:WaistCondition} are satisfied. 
Correspondingly, in the following we choose $\theta_{\ast} = 450~\text{\textmu{}rad}$ and $w_{\mathsf{x}} = 2~\text{\textmu{}m}$. 
The remaining parameters for the x-ray and optical lasers are given in \tabref{tab:Parameters}, where we choose $\fnum = 2$ focusing for the optical lasers.

\begin{figure}[t!!]
    \includegraphics[width=0.45\textwidth,trim={0.0cm 0.0cm 0.0cm 0.0cm},clip=true]{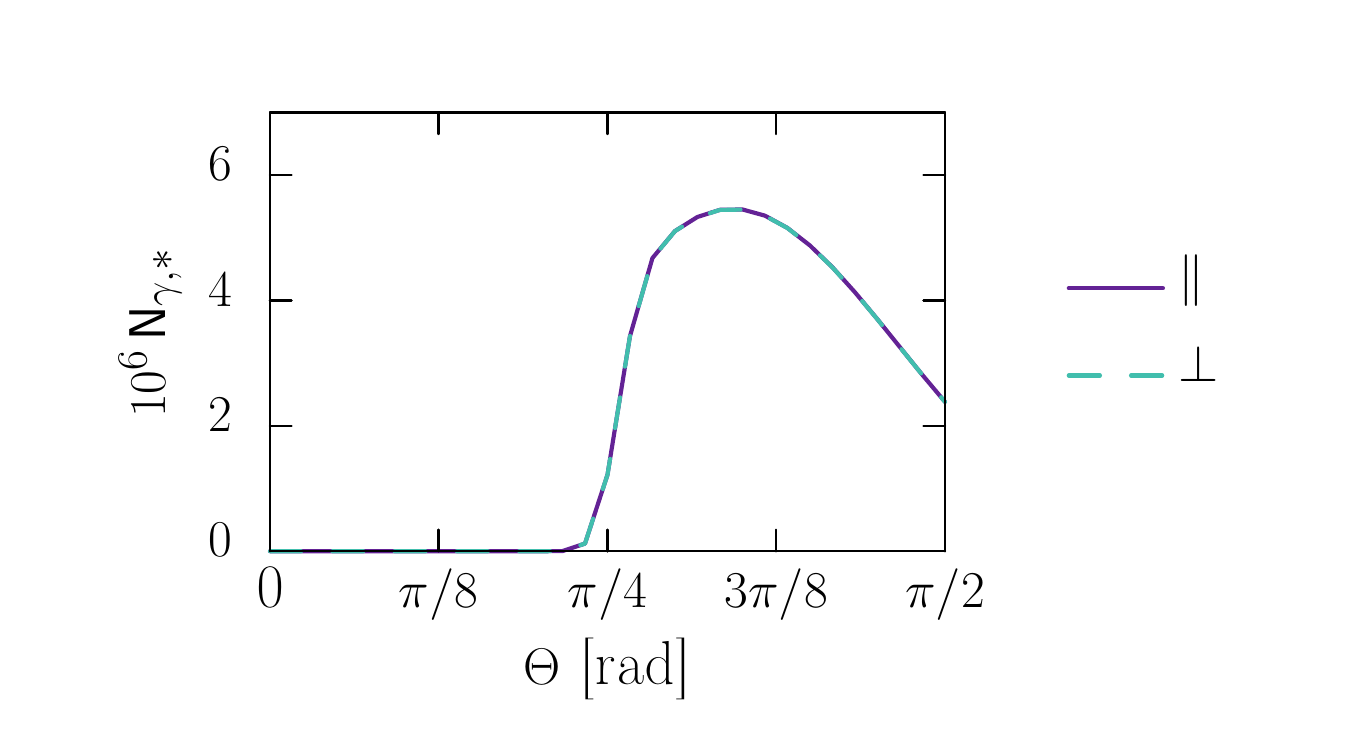}
    \caption{\label{fig:ModesNumberAngleFuture} 
        Number of signal photons vs collision angle $\Theta$ from the collision of EuXFEL self-seeded $\omega_{\tsf{x}} = 10~\mathrm{keV}$ photons with two future laser pulses.
        X-ray pulses are focused to $w_{\tsf{x}} = 4~\text{\textmu{}m}$ and optical pulses are focused with $\fnum = 2$.
        X-ray and optical pulses have polarization angle $\psi = \psi_{\star}$ [c.f. \eqnref{eqn:Psistar}], at which the numbers of parallel and perpendicular photons are equal.
        Plotted are the numbers of parallel $\tsf{N}_{\gamma,\ast}^{\parallel}$ (purple solid line) and perpendicular $\tsf{N}_{\gamma,\ast}^{\perp}$ (blue dashed line) signal photons with emission angles $\theta_{x} > 450~\text{\textmu{}rad}$.
    }
\end{figure}

Consider now the birefringent signal. To determine the fundamental constants using \eqnref{eqn:NumberTotalApproxNLOSimple} for a fixed polarization angle of the x-ray and optical beams, the number of both parallel polarized and perpendicular polarized photons must be measured. We saw from \figref{fig:Prefactors} that as the optical and x-ray polarization plane angle, $\psi$, is varied, the number of photons scattered into the perpendicular polarization mode is maximized where the number in the parallel mode vanishes and vice versa. To determine a suitable value for $\psi$, we pick the value at which the ratio $\tsf{N}_{\gamma}^{\perp}/\tsf{N}_{\gamma}^{\parallel} \to 1$. From \eqnref{eqn:RatioNLO} this occurs when
\begin{align}\label{eqn:Psistar}
    \psi = \psi_{\star} 
    =
    \begin{cases}
        \frac{
            \tan^{-1}\big(\frac{3 \cc}{\cd}\big)
        }{2} \approx 30.8^{\circ}\\
        \frac{
            \pi - \tan^{-1}\big(\frac{3 \cc}{\cd}\big)
        }{2}
        \approx 59.2^{\circ}
        \,,
    \end{cases}
\end{align}
where the approximated values are those obtained using \eqnref{eqn:Coefficients34}. 
For other choices of $\psi$, one of $\tsf{N}_{\gamma}^{\perp}$ or $\tsf{N}_{\gamma}^{\parallel}$ will be smaller, making measurement more difficult.
The choice of $\psi = \psi_{\star} = \tan^{-1}(3 \cc/\cd)/2$ is selected for \figref{fig:ModesNumberAngleFuture}, which shows the number of signal photons in each polarization mode with scattering angles $\theta_{x} > \theta_{\ast}$.
As expected from our choice of $\psi = \psi_{\star}$, we find that the number of parallel and perpendicular polarized photons coincide, peaking around $\Theta \approx 63^{\circ}$ with the value $\tsf{N}_{\gamma,\ast}^{\parallel} = \tsf{N}_{\gamma,\ast}^{\perp} \approx 5.5 \times 10^{-6}$.
Following the same analysis as in the preceding section of estimating the background per shot and taking into account the x-ray optics, we find
\begin{equation}\label{eqn:BirefringentNumbersFuture}
    \begin{tabular}{c c c c}
        $\mathsf{N}_{\gamma,\ast}^{\parallel}$ 
        & 
        $\mathsf{N}_{\gamma}^{\mathsf{bg}\parallel}$ 
        & 
        $\mathsf{N}_{\gamma,\ast}^{\perp}$ 
        & 
        $\mathsf{N}_{\gamma}^{\mathsf{bg}\perp}$ 
        \\
        \hline
        $1.4 \times 10^{-6}$
        &
        $1.3 \times 10^{-7}$
        &
        $1.4 \times 10^{-6}$
        &
        $1.5 \times 10^{-12}$
    \end{tabular}
\end{equation}
which amounts to $\mathsf{N}_{\mathsf{shots}}^{5\sigma,\parallel} \gtrsim 7.2 \times 10^{6}$ and $\mathsf{N}_{\mathsf{shots}}^{5\sigma,\perp} \gtrsim 7.1 \times 10^{5}$.
While the number of shots is perhaps prohibitively large for a birefringent measurement to be feasible using the proposed parameters, particularly since the high-pulse-energy requirement of the optical pulses would likely mean a low repetition rate, we emphasise that the number of signal photons is directly proportional to the number of XFEL photons $\mathsf{N}_{\mathsf{x}}$.
We have considered here a next-generation high-power laser combined with an XFEL with currently available technology.
Thus, it is possible that advances in XFEL technology could push the birefringent signal to a viable level for detection in an experiment.

\begin{figure}
    \includegraphics[width=0.45\textwidth,trim={0.0cm 0.0cm 0.0cm 0.0cm},clip=true]{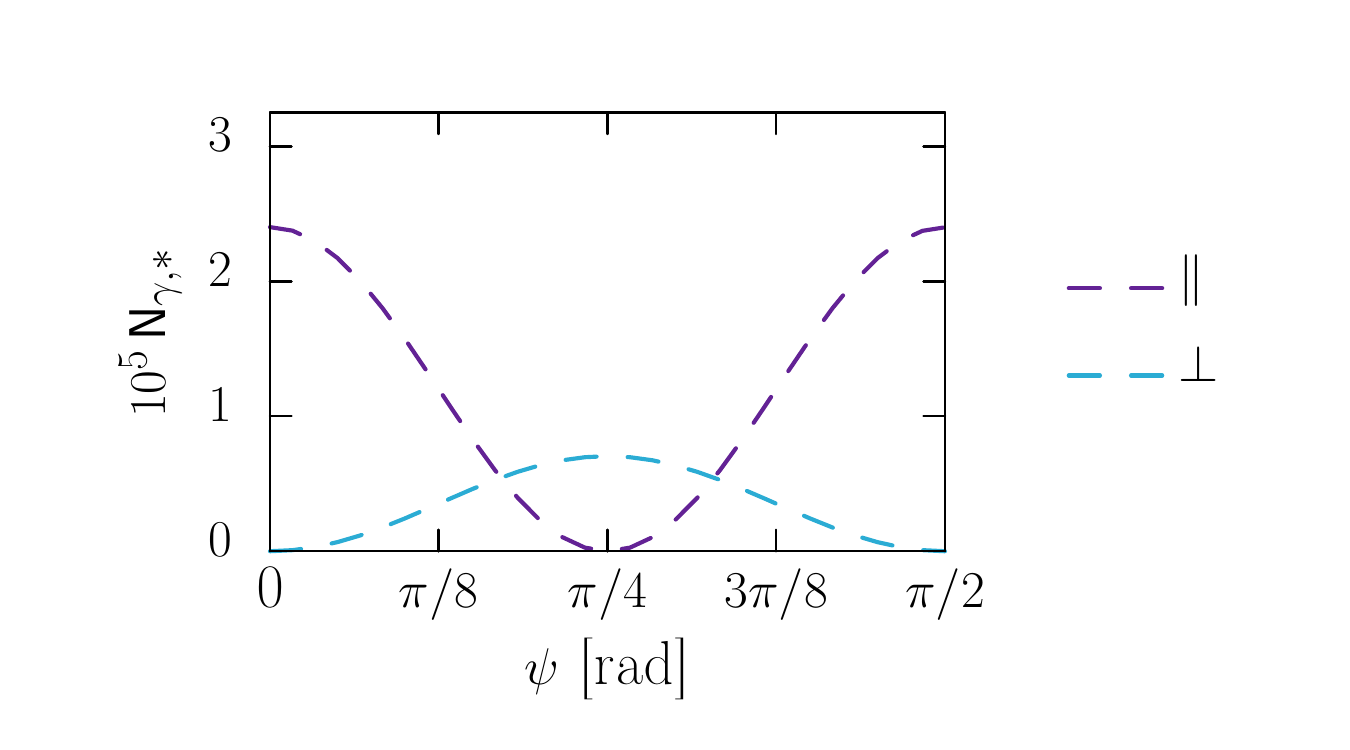}
    \caption{\label{fig:ModesNumberPolarisationFuture} 
        Dependence of the signal photons $\tsf{N}_{\gamma,\ast}^{\parallel,\perp}$ on the polarization $\psi$ for fixed collision angle $\Theta = 63^{\circ}$.
        The x-ray and optical pulse parameters are as in \figref{fig:ModesNumberAngleFuture}.
        Plotted are the number of photons $\tsf{N}_{\gamma}^{\parallel}$ scattered into the parallel state  (purple solid line), and the number $\tsf{N}_{\gamma}^{\perp}$ scattered into the perpendicular state (blue dashed line), if an angular cut is applied which only accepts photons with emission angles $\theta_{x} > 450~\text{\textmu{}rad}$.
    }
\end{figure}

\begin{figure}
    \includegraphics[width=0.45\textwidth,trim={0.0cm 0.0cm 0.0cm 0.0cm},clip=true]{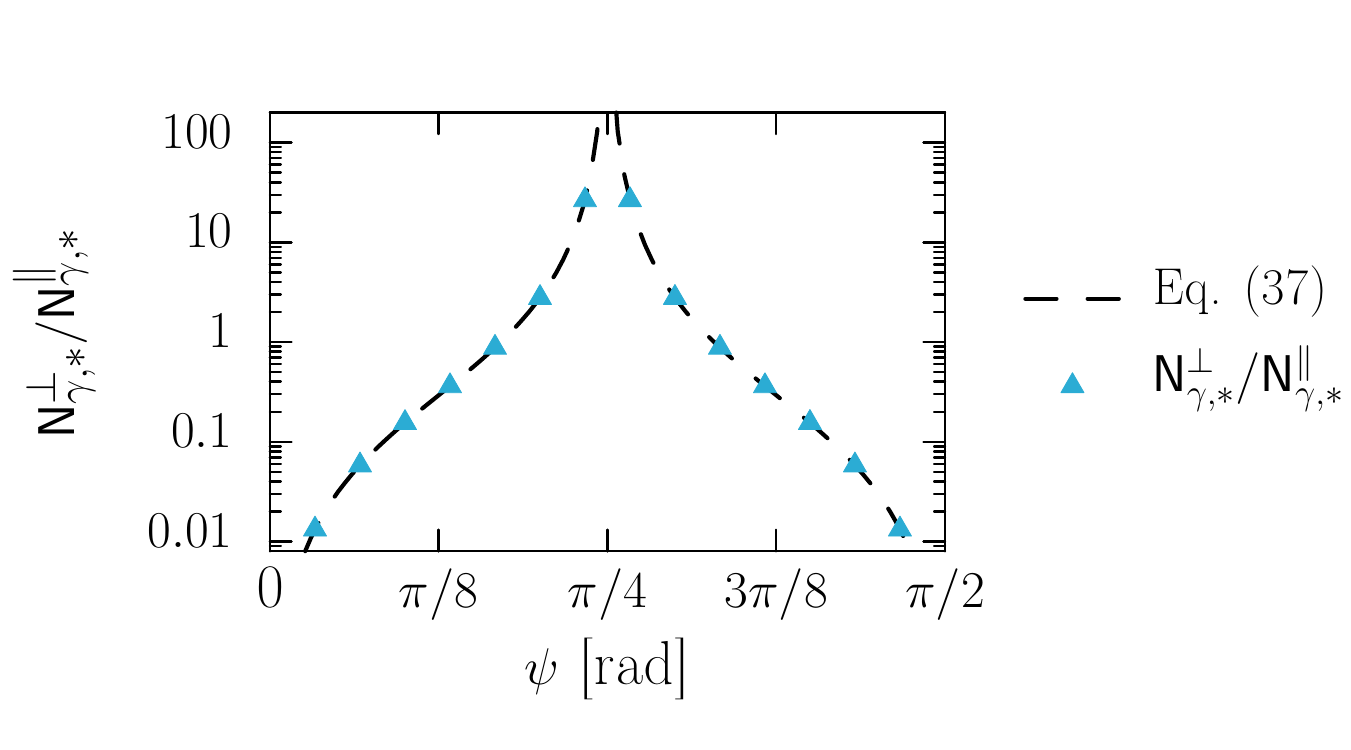}
    \caption{\label{fig:ModesRatioPolarisationFuture} 
        Ratio of the number of photons in each polarization mode $\tsf{N}_{\gamma}^{\perp}/\tsf{N}_{\gamma}^{\parallel}$ vs polarization $\psi$ for fixed collision angle $\Theta = 63^{\circ}$.
        The x-ray and optical pulse parameters as in \figref{fig:ModesNumberAngleFuture}.
        The black dashed line shows the analytical result from \eqnref{eqn:RatioNLO} with $\cc$ and $\cd$ given by \eqnref{eqn:Coefficients34},
        and the blue triangles show the numerically evaluated ratio of signal photons with $\theta_{\gamma} > \theta_{\ast}$, $\tsf{N}_{\gamma,\ast}^{\perp}/\tsf{N}_{\gamma,\ast}^{\parallel}$.
    }
\end{figure}

Before moving on to consider the polarization-insensitive case we consider how the ratio $\tsf{N}_{\gamma}^{\perp}/\tsf{N}_{\gamma}^{\parallel}$ varies for the NLO contribution.
Choosing the collision angle $\Theta = 63^{\circ}$ which maximizes the birefringent signal of photons with $\theta_{x} > 450~\text{\textmu{}rad}$ in \figref{fig:ModesNumberAngleFuture}, we plot in \figref{fig:ModesNumberPolarisationFuture} how the number of signal photons in each polarization mode varies with the angle $\psi$ and in \figref{fig:ModesRatioPolarisationFuture} compare their ratio against the analytical approximation \eqnref{eqn:RatioNLO}, using \eqnref{eqn:Coefficients34}.
We find perfect agreement between the two cases.

\subsection{Polarization-insensitive measurements \label{sec:Scattering}}

\subsubsection{Currently available  technology: four-photon scattering}\label{sec:CurrentpolarizationInsensitive}

\begin{figure}
    \includegraphics[width=0.45\textwidth,trim={0.0cm 0.0cm 0.0cm 0.0cm},clip=true]{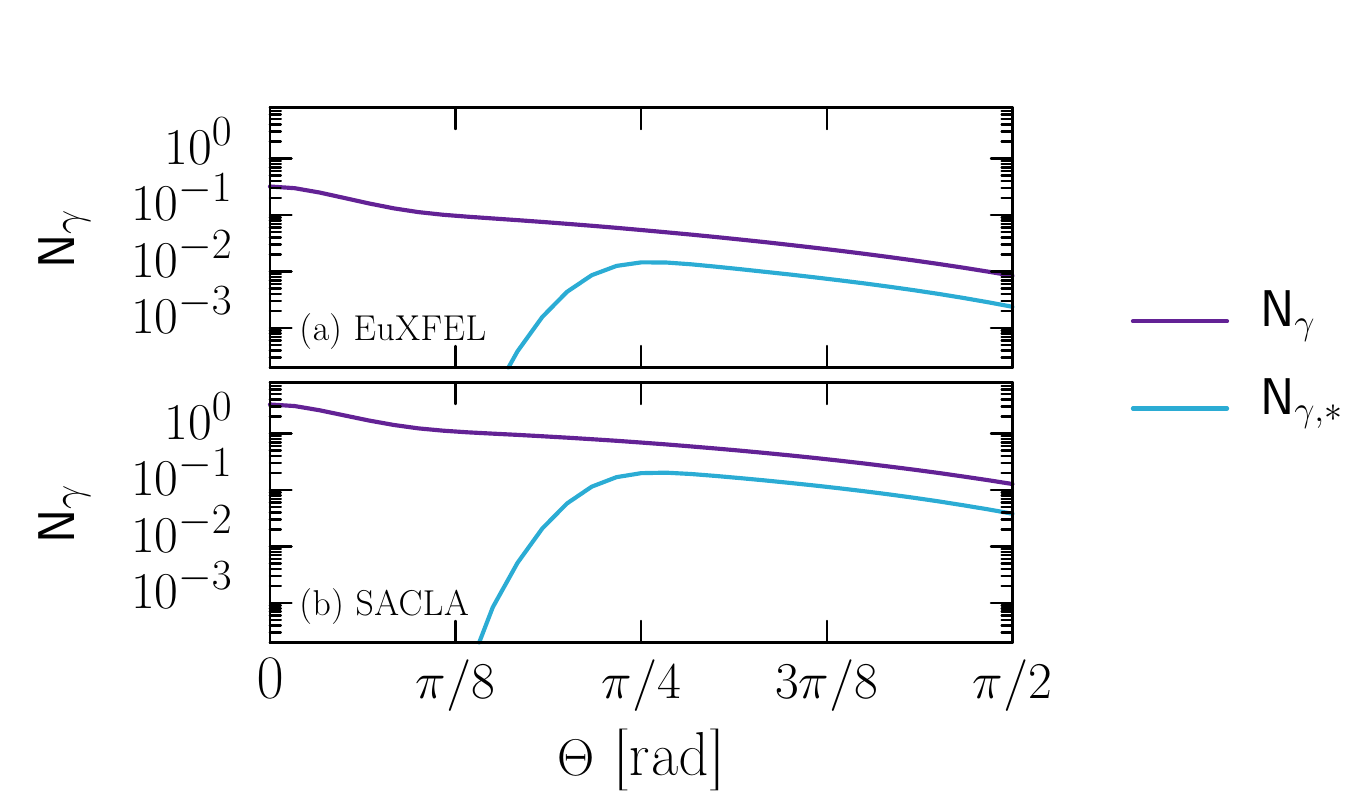}
    \caption{\label{fig:TotalNumberAngleCurrent} 
    Number of signal photons vs collision angle $\Theta$ in a polarization-insensitive measurement using (a) EuXFEL and (b) SACLA SASE parameters.
    Optical pulses are focused with $\fnum = 1$ focusing, x-ray pulses are focused to $w_{\tsf{x}} = 4~\text{\textmu{}m}$, and relative polarization angle $\Delta\psi = \pi/2$. 
    Plotted is the total number of photons $\tsf{N}_{\gamma}$ (purple solid line) and the equivalent photon counts if an angular cut is applied which only accepts photons with emission angles $\theta_{x} > 200~\text{\textmu{}rad}$ (blue solid line).
    }
\end{figure}

We now consider measurement of the unpolarized process.
One advantage of a polarization-insensitive measurement is that the SASE mode of the XFEL can be used, which increases the number of photons per bunch (see \tabref{tab:Parameters}).
From Eqs. \eqref{eqn:NumberTotalApproxLO} and \eqref{eqn:Coefficients12} it is clear that the total number of signal photons is maximized when the difference in polarization angle $\Delta\psi = \pi/2$.
In this case the number of perpendicular polarized photons drops to zero [see \eqnref{eqn:NumberTotalApproxLO}].
The dependence of the total number of signal photons versus the collision angle $\Theta$ is shown in \figref{fig:TotalNumberAngleCurrent} for $\fnum = 1$ focusing with EuXFEL [\figref{eqn:NumberTotalApproxLO}(a)] and SACLA [\figref{eqn:NumberTotalApproxLO}(b)] SASE parameters.
The purple solid lines show the total number of signal photons with all scattering angles, $\mathsf{N}_{\gamma}$, i.e. including those in the central peak (see \figref{fig:AngleCurrent}).
Since the signal photons are purely parallel polarized, the photons in the central peak will of course be indistinguishable from the background probe XFEL photons, and so the blue solid lines show the total number of signal photons with angles $\theta_{x} > 200~\text{\textmu{}rad}$, $\mathsf{N}_{\gamma,\ast}$.
This peaks at the collision angle $\Theta \approx 45^{\circ}$ with $\mathsf{N}_{\gamma,\ast} \approx 1.5 \times 10^{-2}$ for EuXFEL parameters, and $\Theta \approx 48^{\circ}$ with $\mathsf{N}_{\gamma,\ast} \approx 2.0 \times 10^{-1}$ for SACLA.
The combination of the increase in the number of probe photons, $\mathsf{N}_{\mathsf{x}}$, and the use of the polarization difference $\Delta\psi = \pi/2$ leads to almost an order of magnitude increase in the number of signal photons per shot with comparison to the polarization-sensitive measurements (\figref{fig:ModesNumberAngleCurrent}).

\begin{figure}
    \includegraphics[width=0.45\textwidth,trim={0.0cm 0.0cm 0.0cm 0.0cm},clip=true]{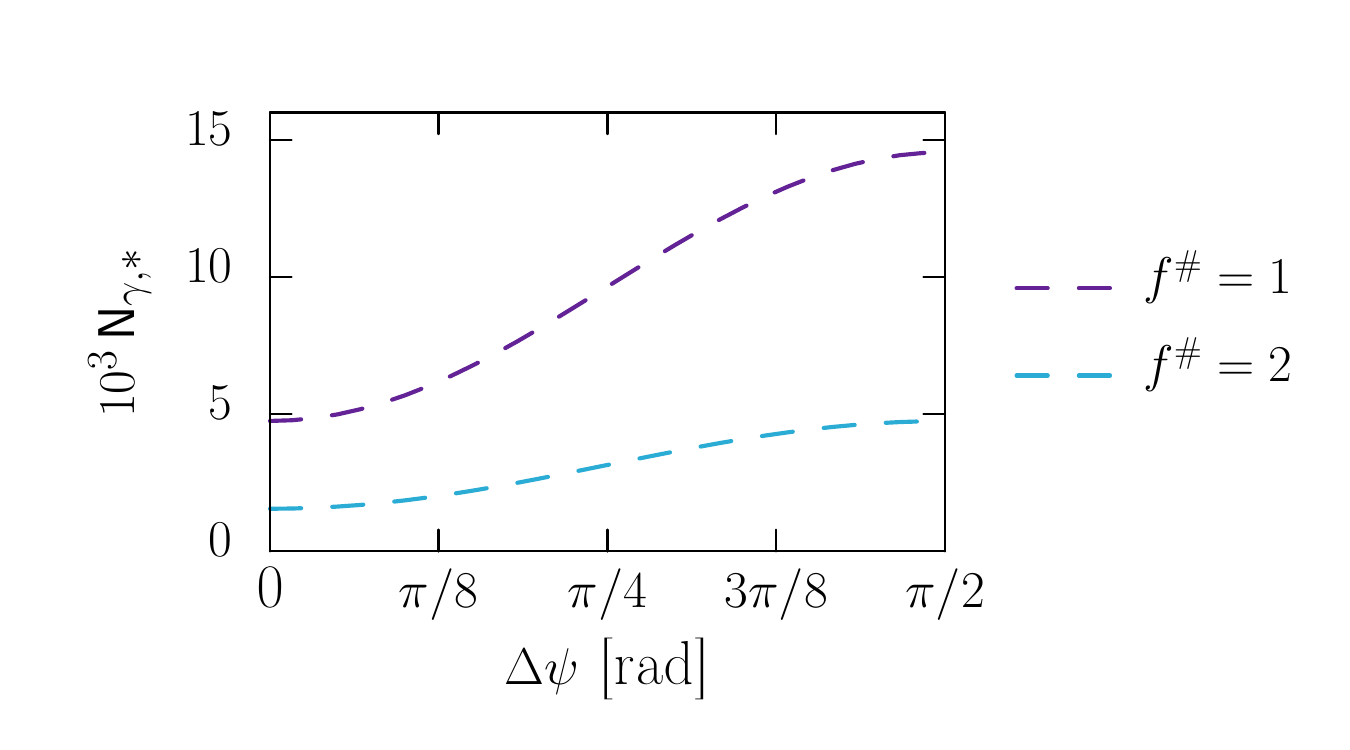}
    \caption{\label{fig:TotalNumberPolarisationCurrent} 
        Dependence of the signal photons $\tsf{N}_{\gamma,\ast}$ on the relative polarization $\Delta\psi$ for fixed collision angle $\Theta = 45^{\circ}$ using EuXFEL SASE parameters.
        Plotted is the total number of photons $\tsf{N}_{\gamma,\ast}$  detected if an angular cut is applied which only accepts photons with emission angles $\theta_{x} > 200~\text{\textmu{}rad}$ for $\fnum = 1$ (purple) and $\fnum = 2$ (blue) focusing of the optical laser.
    }
\end{figure}

To enable the determination of the experimental values of the four-photon low-energy constants $\ca$ and $\cb$ it is necessary to perform at least two independent measurements.
For the polarization-sensitive measurement of \secref{sec:Current} these were the number of photons scattered into the parallel and perpendicular polarization modes.
For a polarization-insensitive measurement, where only the total number of signal photons is obtained, the two independent observables must come from measurements at different choices of the relative angle of the XFEL and optical polarization planes, $\Delta\psi$. How the number of signal photons varies with $\Delta\psi$ for fixed collision angle $\Theta = 45^{\circ}$ is shown in \figref{fig:TotalNumberPolarisationCurrent} for EuXFEL parameters.
Unlike in the polarization-sensitive measurement, where the difference in the number of parallel and perpendicular signal photons per shot is over an order of magnitude, the difference between measurements at $\Delta\psi = 0$ and $\Delta\psi = \pi/2$ is only around a factor of $3$.
This means that each measurement will require a comparable number of shots to reach a particular statistical significance and, coupled with the higher number of signal photons per shot, the total required shots to determine $\ca$ and $\cb$ will be less than in the polarization-sensitive case.

\begin{table*}
    \centering
    {\setlength{\tabcolsep}{1em}
    \begin{tabular}{l c c c c c c }
        \hhline{=======}
        Case
        &
        $\fnum$
        &
        $\mathsf{N}_{\gamma,\ast}(\pi/4)$ 
        & 
        $\mathsf{N}_{\gamma,\ast}(\pi/2)$ 
        & 
        $\mathsf{N}_{\gamma}^{\mathsf{bg}}$ 
        &
        $\mathsf{N}_{\mathsf{shots}}^{3\sigma}$ 
        &
        $\mathsf{N}_{\mathsf{shots}}^{5\sigma}$ 
        \\
        \hline
        EuXFEL
        &
        1
        &
        $2.4 \times 10^{-3}$
        &
        $3.6 \times 10^{-3}$
        &
        $6.3 \times 10^{-7}$
        &
        $4.2 \times 10^{2}$
        &
        $1.2 \times 10^{3}$
        \\
        EuXFEL
        &
        2
        &
        $7.9 \times 10^{-4}$
        &
        $1.2 \times 10^{-3}$
        &
        $6.3 \times 10^{-7}$
        &
        $1.5 \times 10^{3}$
        &
        $4.2 \times 10^{3}$
        \\
        SACLA
        &
        1
        &
        $3.3 \times 10^{-2}$
        &
        $5.0 \times 10^{-2}$
        &
        $2.5 \times 10^{-7}$
        &
        $2.1 \times 10^{1}$
        &
        $5.8 \times 10^{1}$
        \\
        SACLA
        &
        2
        &
        $8.4 \times 10^{-3}$
        &
        $1.3 \times 10^{-2}$
        &
        $2.5 \times 10^{-7}$
        &
        $9.3 \times 10^{1}$
        &
        $2.6 \times 10^{2}$
        \\
        \hhline{=======}
    \end{tabular}
    }
    \caption{\label{tab:TotalNumbers} 
        Estimated number of signal photons in a polarization-insensitive measurement, $\mathsf{N}_{\gamma,\ast}(\Delta\psi)$, for different polarization differences $\Delta\psi$, along with the associated background $\mathsf{N}_{\gamma}^{\mathsf{bg}}$.
        Estimates account for the use of two standard beryllium lenses with 50\% transmission to focus the probe and scattered beams.
        Also shown is the estimated lower bound on the number of shots  required for $3\sigma$ and $5\sigma$ statistical significance, $\mathsf{N}_{\mathsf{shots}}^{\mathsf{n}\sigma} = \mathsf{N}_{\mathsf{shots}}^{\mathsf{n}\sigma,\pi/4} + \mathsf{N}_{\mathsf{shots}}^{\mathsf{n}\sigma,\pi/2}$, calculated assuming an optimal collision.
    }
\end{table*}

\begin{figure}
    \includegraphics[width=0.45\textwidth,trim={0.25cm 0.0cm 0.0cm 0.5cm},clip=true]{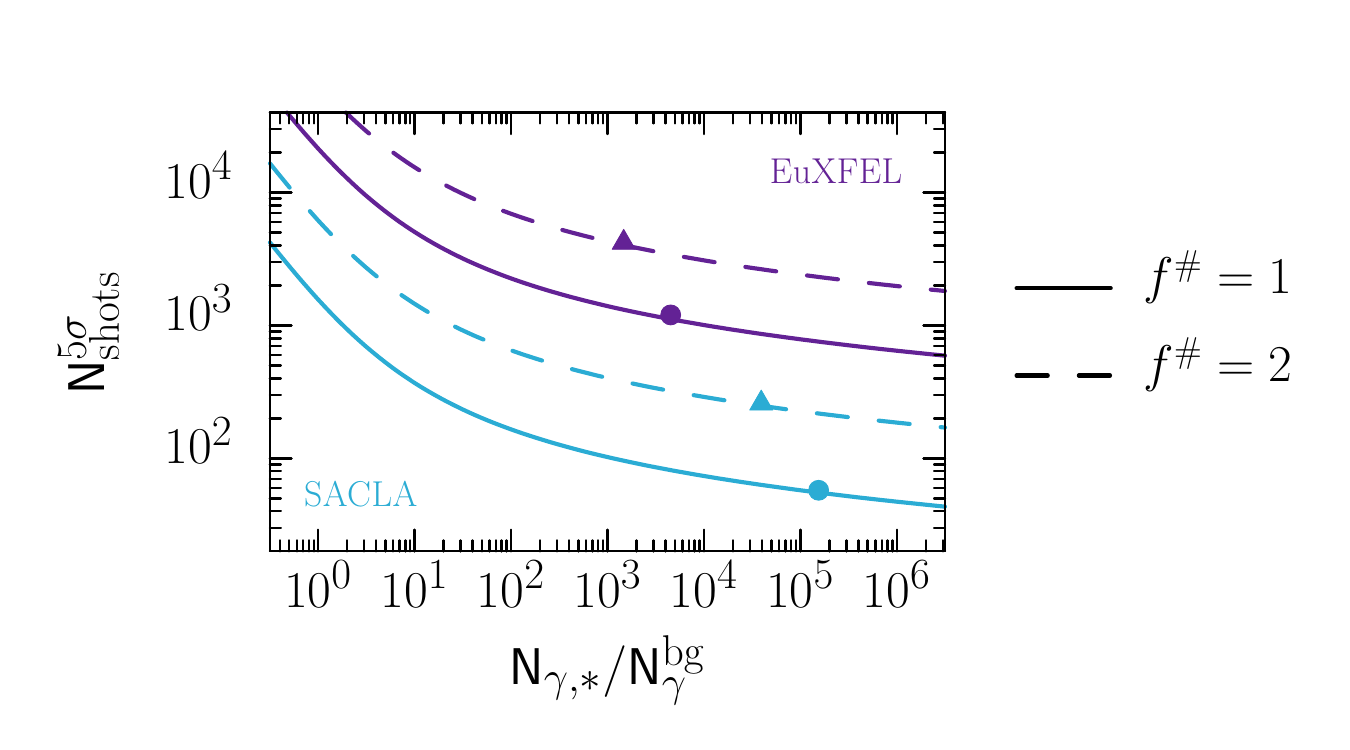}
    \caption{\label{fig:TotalShots} 
        Estimated lower bound on the number of shots  required for $5\sigma$ statistical significance, $\mathsf{N}_{\mathsf{shots}}^{\mathsf{n}\sigma} = \mathsf{N}_{\mathsf{shots}}^{\mathsf{n}\sigma,\pi/4} + \mathsf{N}_{\mathsf{shots}}^{\mathsf{n}\sigma,\pi/2}$, as a function of the signal-to-background ratio $\mathsf{N}_{\gamma,\ast}/\mathsf{N}_{\gamma}^{\mathsf{bg}}$.
        X-ray and optical pulse parameters as in \figref{fig:TotalNumberAngleCurrent} with $\Theta = 45^{\circ}$.
        Estimations of number of signal photons held fixed and given in \tabref{tab:TotalNumbers}.
        \emph{Purple solid: } $\fnum = 1$ (EuXFEL).
        \emph{Purple dashed: } $\fnum = 2$ (EuXFEL). 
        \emph{Blue solid: } $\fnum = 1$ (SACLA).
        \emph{Blue dashed: } $\fnum = 2$ (SACLA).
        The data points correspond to the background estimations in \tabref{tab:TotalNumbers}. 
    }
\end{figure}

We again wish to estimate the minimum number of shots required to obtain a $5\sigma$ statistical significance, $\mathsf{N}_{\mathsf{shots}}^{5\sigma}$.
With the background from the x-ray divergence $\mathsf{N}_{\mathsf{x}}^{\mathsf{bg}} \approx 0$ and the polarization purity playing no role, the main source of background in the polarization-insensitive measurement will be from Compton scattering, $\mathsf{N}_{\gamma}^{\mathsf{bg}} \approx \mathsf{N}_{\mathsf{C}}^{\mathsf{bg}}$, which is estimated as $\mathsf{N}_{\mathsf{C}}^{\mathsf{bg}} \approx 2.5 \times 10^{-6}$ for EuXFEL parameters and $\mathsf{N}_{\mathsf{C}}^{\mathsf{bg}} \approx 1.0 \times 10^{-6}$ for SACLA parameters.
Accounting for the reduction of both the signal and background by the x-ray optics, the number of signal photons for different $\Delta\psi$ measurements and optical focusing are presented in \tabref{tab:TotalNumbers}.
Taking the independent observables to be the measurements at $\Delta\psi = (\pi/4,\pi/2)$ at $\fnum = 1$ optical focusing, these estimates correspond to $\mathsf{N}_{\mathsf{shots}}^{5\sigma,\pi/4} \gtrsim 7.1 \times 10^{2}$ and $\mathsf{N}_{\mathsf{shots}}^{5\sigma,\pi/2} \gtrsim 4.5 \times 10^{2}$  for EuXFEL parameters, and 
$\mathsf{N}_{\mathsf{shots}}^{5\sigma,\pi/4} \gtrsim 3.5 \times 10^{1}$ and
$\mathsf{N}_{\mathsf{shots}}^{5\sigma,\pi/2} \gtrsim 2.2 \times 10^{1}$  for SACLA parameters.
The total number of shots required to measure both observables will be the sum $\mathsf{N}_{\mathsf{shots}}^{5\sigma} \gtrsim \mathsf{N}_{\mathsf{shots}}^{5\sigma,\pi/4} + \mathsf{N}_{\mathsf{shots}}^{5\sigma,\pi/2}$, which are also shown in \tabref{tab:TotalNumbers}.
Comparing with \tabref{tab:BirefringentNumbers}, we see that over an order of magnitude fewer shots would be required to determine the experimental values of the fundamental low energy constants $\ca$ and $\cb$ to $5\sigma$ statistical significance. 
This is also seen in \figref{fig:TotalShots}, which shows how the lower bound of the number of shots required for $5\sigma$ significance in the polarization-insensitive measurement varies with the background using the estimated number of signal photons from \tabref{tab:TotalNumbers} with $\mathsf{N}_{\mathsf{shots}}^{5\sigma} \gtrsim \mathsf{N}_{\mathsf{shots}}^{5\sigma,\pi/4} + \mathsf{N}_{\mathsf{shots}}^{5\sigma,\pi/2}$ and \cref{eqn:Shots}.

\begin{figure}
    \includegraphics[width=0.45\textwidth,trim={0.0cm 0.0cm 0.0cm 0.0cm},clip=true]{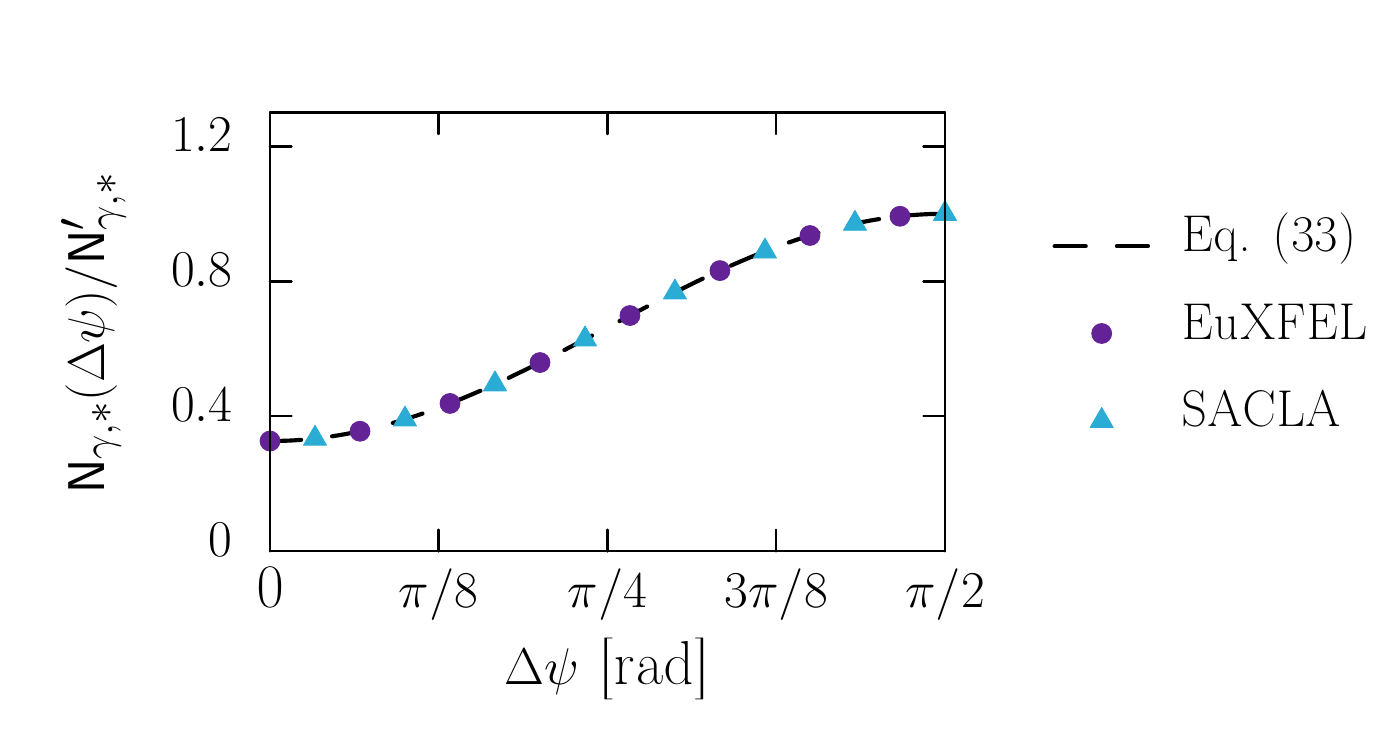}
    \caption{\label{fig:TotalRatioPolarisationCurrent} 
        Ratio of the total number of scattered photons at different relative polarizations, $\mathsf{N}_{\gamma}(\Delta\psi)/\mathsf{N}_{\gamma}(\Delta\psi^{\prime})$.
        The reference relative polarization is chosen as $\Delta\psi^{\prime} = \pi/2$.
        The x-ray and optical pulse parameters are as in \figref{fig:TotalNumberAngleCurrent}.
        The black dashed line shows the ratio \eqnref{eqn:NRatioLO} for $\Delta\psi^{\prime} = \pi/2$,
        purple circles show the EuXFEL SASE parameters,
        and the blue triangles show the SACLA SASE parameters.
    }
\end{figure}

When considering the birefringent signal, it was possible to demonstrate through \figref{fig:ModesRatioPolarisationCurrent} that the ratio of the number of perpendicular to parallel polarized signal photons follow simple analytical relationships which depend only on the fundamental low-energy constants of the LO process and the polarization difference [\eqnref{eqn:RatioLO}].
We can demonstrate an analogous relationship in the polarization-insensitive measurement by instead considering the ratio of the total number of signal photons due to different polarization differences, $\Delta\psi \ne \Delta\psi^{\prime}$ [see \eqnref{eqn:NRatioLO}].
This is shown in \figref{fig:TotalRatioPolarisationCurrent} for a fixed collision angle of $\Theta = 30^{\circ}$ for $\fnum = 1$ optical focusing.
We take as a reference value the number of signal photons at $\Delta\psi^{\prime} = \pi/2$, $\mathsf{N}_{\gamma,\ast}^{\prime} \equiv \mathsf{N}_{\gamma,\ast}(\pi/2)$, and use this to normalize the number of photons as a function of $\Delta\psi$, $\mathsf{N}_{\gamma,\ast}(\Delta\psi)$.
This normalization ensures that all of the space-time dependent factors cancel, and we see that the simple analytical ratio \eqnref{eqn:NRatioLO} agrees with the numerically evaluated data for both EuXFEL (purple circles) and SACLA (blue triangles) SASE parameters.

\subsubsection{Next generation technology: six-photon scattering}\label{sec:FuturepolarizationInsensitive}

We now consider a measurement of unpolarized photon-photon scattering with an upgraded optical laser \tabref{tab:Parameters} as in \secref{sec:FuturepolarizationSensitive} (with $w_{\mathsf{x}} = 2~\text{\textmu{}m}$, $\fnum = 2$, and $\theta_{\ast} = 450~\text{\textmu{}rad}$). The polarization $\psi$ is chosen to optimize the number of scattered photons: $\psi_{\tsf{x}} = \psi_{0} = \psi$.
At fixed $\psi = 0$, which maximizes the total number of signal photons, the use of SASE parameters increases the number of photons with $\theta_{\gamma} > \theta_{\ast}$ to $\mathsf{N}_{\gamma,\ast} \approx 1.2 \times 10^{-4}$ at the peak collision angle $\Theta = 63^{\circ}$.
While still small, this is a large improvement over the polarization-sensitive measurement outlined in \secref{sec:FuturepolarizationSensitive}.

Just as in the LO case, we need two independent measurements to be able to experimentally determine the values of the six-photon low-energy constants, theoretically predicted as in \eqnref{eqn:Coefficients34}.
The two observables are chosen as the total number of signal photons at two different values of the polarization parameter $\psi$.
We choose the collision angle $\Theta = 63^{\circ}$ around which the number of signal photons has a maximum.
In this case we find, including the effect of x-ray focusing optics,
\begin{equation}\label{eqn:TotalNumbersFuture}
    \begin{tabular}{c c c}
        $\mathsf{N}_{\gamma,\ast}(\pi/4)$
        &
        $\mathsf{N}_{\gamma,\ast}(\pi/2)$
        &
        $\mathsf{N}_{\gamma}^{\mathsf{bg}}$
        \\
        \hline
        $3.0 \times 10^{-5}$
        &
        $8.8 \times 10^{-6}$
        &
        $6.3 \times 10^{-7}$
    \end{tabular}
\end{equation}
with the estimated backgrounds being the same as in the preceding section.
We estimate that $\mathsf{N}_{\mathsf{shots}}^{5\sigma} \gtrsim 1.1 \times 10^{6}$ shots would be required to determine the six-photon low-energy constants to $5\sigma$ significance.

\section{Summary \label{sec:Summary}}

Real photon-photon scattering has yet to be experimentally verified. 
Measuring either the unpolarized or the polarized case (vacuum birefringence) would allow confirmation of an outstanding prediction from QED for the magnitude of fundamental low-energy constants that govern the effective nonlinear coupling of photons with one other. 
Measurement of the effective photon-photon coupling could be used to place bounds on new physics beyond the standard model \cite{Ahlers:2007qf,Gies:2008wv,Villalba-Chavez:2016hxw,Villalba-Chavez:2016hht,Fouche:2016qqj,Beyer:2020dag,Ejlli:2020yhk,Evans:2023jpr} and act as a gateway to harnessing the nonlinear vacuum for more exotic applications such as vacuum high harmonic generation and self-focusing.

We have shown that there are several advantages in using a planar three-beam setup to measure photon-photon scattering.
The kinematics allow for photons to be scattered into Bragg side peaks in the detector plane, thereby providing spatial separation of the signal from the large photon background and significantly increasing the signal-to-background ratio compared to the more conventional two-beam setup. 
In the planar three-beam setup, (i) the fundamental low-energy QED constants can be determined by measuring numbers of scattered photons, without the need for polarimetry, ii) since the bandwidth of the x-ray beam does not have to be especially low for polarimetry purposes, an XFEL seeding mode such as SASE can be used that increases the photon-photon scattering signal; and
(iii) higher orders of photon-photon scattering lead to further Bragg side peaks that correlate transverse detector position with expansion order of the effective photon-photon interaction. 
This in principle allows the determination of fundamental low-energy QED constants for higher-dimensional terms that lead to, e.g., six-photon scattering. 
These considerations further emphasise the point that configurations of lasers that are beyond plane-waves can support processes that are otherwise kinematically suppressed in the plane-wave case.

Specifically, we calculated the number of scattered photons if an XFEL provides the probe field and two high power optical beams provide the pump field. 
Two different parameter sets were considered.
First, currently available technology was shown to be sufficient to measure four-photon scattering; parameters were chosen from those available at the EuXFEL, which will be used in the proposed BIREF@HIBEF experiment \cite{Schlenvoigt:2016jrd,Ahmadiniaz:2024xob}, and those available at SACLA.
We found that the number of required shots to obtain a $5\sigma$ significance value using the three-beam scenario  (c.f.~Tables \ref{tab:BirefringentNumbers} and \ref{tab:TotalNumbers}) could be competitive with, and/or complementary to, other approaches such as the dark-field method~\cite{Karbstein:2022uwf,Ahmadiniaz:2024xob}.
Secondly, we considered future technology to measure six-photon scattering with the EuXFEL and an optical laser capable of reaching intensities of approximately $10^{24}\text{-}10^{25}~\mathrm{Wcm^{-2}}$, such as may be achievable at the Station of Extreme Light \cite{Shen:2018lbq}. 
Alternatively, future parameters to increase the six-photon signal could involve increasing the brilliance of the XFEL beam or the frequency of collisions with the optical beams.

\section*{Acknowledgments}

A.J.M. was supported by the project Advanced Research Using High Intensity Laser Produced Photons and Particles, Project No. CZ.02.1.01/0.0/0.0/16\_019/0000789 from European Regional Development Fund.
The authors thank Sergei V.~Bulanov for useful discussions.

\appendix

\section{Beam field profiles \label{app:Beams}}

In this appendix we give more details about the field profiles of the three input beams to our configuration.
The XFEL beam and the two optical beams are all described by Gaussian focused pulses in the paraxial approximation.
The normalized electric-field profiles $E_{i}(x)$ are
\begin{align}\label{eqn:Gaussian}
    E_{i}(x)
    =
    &
    \frac{\mathcal{E}_{i}}{\sqrt{1 + \zeta_{i}^{2}}}
    e^{    
        - 
        (x^{\perp}_{i})^{2}/w_{i}^{2} (1 + \zeta_{i}^{2})
        -
        (k_{i} \cdot x)^{2}/\Phi_{i}^{2}
    }
    \nonumber\\
    &
    \times
    \cos\bigg[
        k_{i} \cdot x + \tan^{-1}\zeta_{i} - 
        \frac{(x^{\perp}_{i})^{2} \zeta_{i}}{w_{i}^{2} (1 + \zeta_{i}^{2})}
    \bigg]
    \,,
\end{align} 
where $\zeta_{i} = \bm{k}_{i} \cdot x / \omega_{i} z_{R,i}$ is the curvature parameter defined in terms of the Rayleigh length $z_{R,i} = \omega w_{i}^{2}/2$, with $w_{i}$ defined as the distance from the focus at which the peak intensity is reduced by a factor $1/e^{2}$, and $\bm{k}_{i}$, the 3-wave vector of $k_{i}^{\mu}$.
The phase duration of the pulses is $\Phi_{i} = \omega_{i} \tilde{\tau}_{i}$, with $\tilde{\tau}_{i}$ the temporal duration at which the pulse intensity falls to $1/e^{2}$ of the peak.
This is related to the FWHM value $\tau_{i}$ by $\tilde{\tau}_{i} \simeq 1.7 \tau_{i}$.
The field strengths $\mathcal{E}_{i}$ can be expressed in terms of the total energy of the pulse as \cite{Karbstein:2017jgh}
\begin{align}\label{eqn:FieldStrengths}
    \mathcal{E}_{\mathsf{x}}^{2}
    =
    &
    8
    \sqrt{\frac{2}{\pi}}
    \frac{\mathsf{N}_{\mathsf{x}} \omega_{\mathsf{x}}}{\Ecr^{2}\pi w_{\mathsf{x}}^{2} \tilde{\tau}_{\mathsf{x}}}
    \,,
    \nonumber\\
    \mathcal{E}_{1}^{2}
    =
    \mathcal{E}_{2}^{2}
    =
    \mathcal{E}_{0}^{2}
    =
    &
    8
    \sqrt{\frac{2}{\pi}}
    \frac{W}{\Ecr^{2}\pi w_{0}^{2} \tilde{\tau}_{0}}
    \,,
\end{align}
where for simplicity we consider both the optical lasers to have the same field strength, waist, and duration, $\mathsf{N}_{\mathsf{x}}$ is the number of photons in the x-ray pulse and $W$ is the energy of the optical pulses.  
The coordinates $x^{\perp}_{i}$ denote the directions transverse to the beam propagation direction.
With the wave vectors outlined in \secref{sec:Geometry}, the transverse coordinates are, for $k_{\mathsf{x}}^{\mu}$,
\begin{align}\label{eqn:k1orth}
    (x^{\perp}_{\mathsf{x}})^{2} 
    =
    x^{2} + y^{2}
    \,;
\end{align}
for $k_{1}^{\mu}(\Theta)$,
\begin{align}\label{eqn:k2orth}
    (x^{\perp}_{1})^{2} 
    =
    (x \cos\theta + z \sin\theta)^{2} + y^{2}
    \,;
\end{align}
and for $k_{2}^{\mu}(\Theta)$,
\begin{align}\label{eqn:k3orth}
    (x^{\perp}_{2})^{2} 
    =
    (x \cos\theta - z \sin\theta)^{2} + y^{2}
    \,.
\end{align}

The signal photons $k_{\gamma}^{\mu}$ are taken to be plane-wave states
\begin{align}\label{eqn:FSignal}
    F_{\gamma}^{\mu\nu}(x)
    =
    \frac{1}{\omega_{\gamma}}
    \big(
        k^{\mu}_{\gamma}
        \varepsilon^{\nu}_{\gamma}
        -
        k^{\nu}_{\gamma}
        \varepsilon^{\mu}_{\gamma}
    \big) 
    E_{\gamma}(x)
    \,,
\end{align}
where
\begin{align}\label{eqn:FSignalE}
    E_{\gamma}(x)
    =
    \frac{1}{\Ecr}
    \sqrt{\frac{\omega_{\gamma}}{2 V}}
    e^{i k_{\gamma} \cdot x}
    \,,
\end{align}
with $V$ a volumetric normalization.

Numerical results presented in \secref{sec:Results} have primarily been obtained using the infinite-Rayleigh-length approximation (IRLA), where $\zeta_{i} \to 0$ in \eqnref{eqn:Gaussian} \cite{King:2010nka,King:2012aw,Gies:2017ygp,King:2018wtn,Karbstein:2021ldz}.
In the planar three-beam collision considered in this paper, the functional dependence of the beam profiles \eqnref{eqn:Gaussian} mean that the temporal and out-of-plane coordinate integrals in \eqnref{eqn:Samplitude} and \eqref{eqn:General} can be performed analytically, as they are Gaussian.
However, with the IRLA the form of the beam profiles simplify considerably,
\begin{align}\label{eqn:GaussianIRLA}
    \lim\limits_{\zeta_{i} \to 0}
    E_{i}(x)
    =
    &
    \mathcal{E}_{i}
    e^{    - 
        (x^{\perp}_{i})^{2}/w_{i}^{2}
        -
        (k_{i} \cdot x)^{2}/\Phi_{i}^{2}
    }
    \cos\big(
        k_{i} \cdot x 
    \big)
    \,,
\end{align}
allowing all of the coordinate integrals in Eqs. \eqref{eqn:Samplitude} and \eqref{eqn:General} to be performed analytically.
\begin{figure}[t!!]
    \includegraphics[width=0.45\textwidth,trim={0.0cm 0.0cm 0.0cm 0.0cm},clip=true]{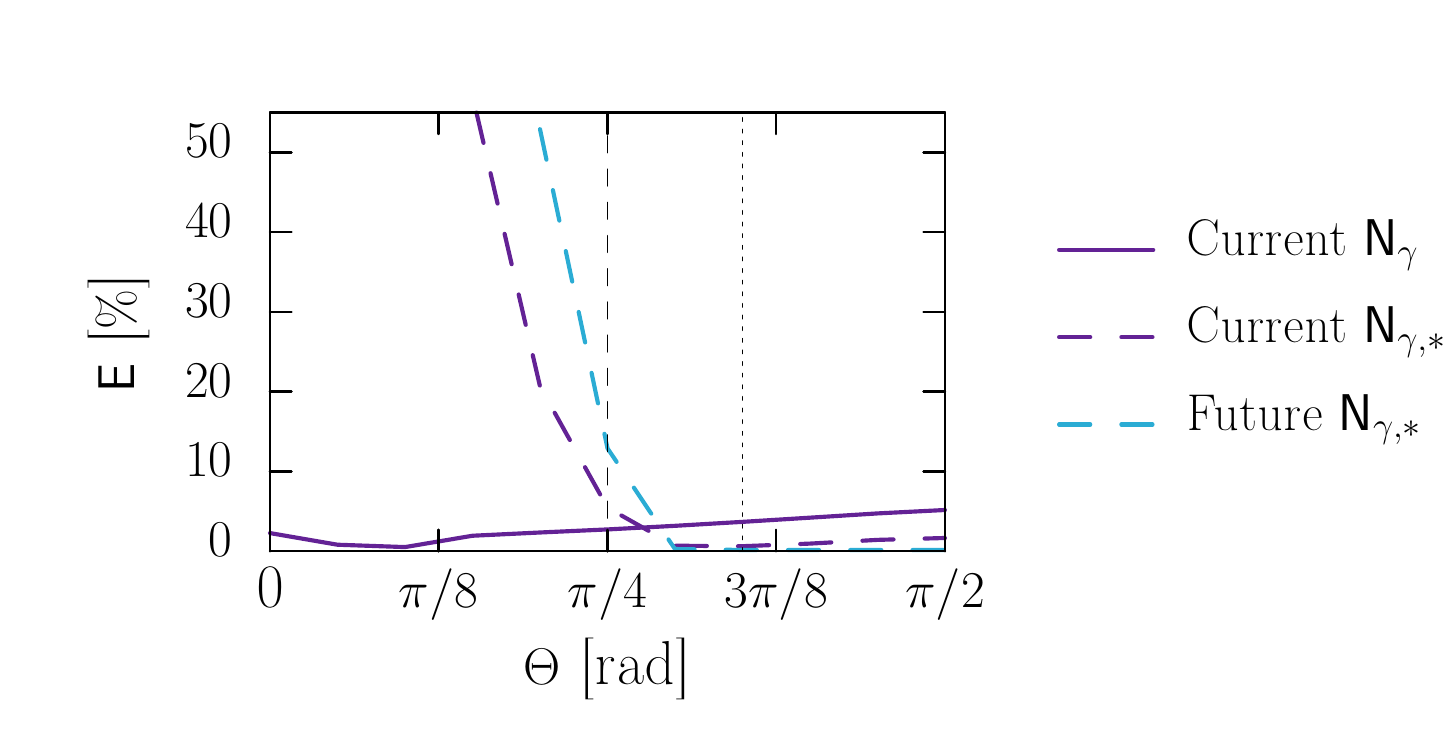}
    \caption{\label{fig:IRLABenchmarking} 
        Relative error $\mathsf{E} = 1 - \mathsf{N}_{\gamma,\ast}^{\mathsf{full}}/\mathsf{N}_{\gamma,\ast}^{\mathsf{IRLA}}$ between the number of signal photons calculated with the full Gaussian pulses in the paraxial approximation, $\mathsf{N}_{\gamma,\ast}^{\mathsf{full}}$, and with the IRLA $\mathsf{N}_{\gamma,\ast}^{\mathsf{IRLA}}$.
        Also shown are the collision angles at which the number of signal photons is maximized for the currently available parameters (vertical dashed) and future parameters (vertical dotted).
    }
\end{figure}
In \figref{fig:IRLABenchmarking} we show the relative error 
\[
    \mathsf{E} 
    = 
    1 
    - 
    \frac{\mathsf{N}_{\gamma,\ast}^{\mathsf{full}}}{\mathsf{N}_{\gamma,\ast}^{\mathsf{IRLA}}}
    \,,
\]
between the number of signal photons calculated with the full Gaussian pulses in the paraxial approximation \eqnref{eqn:Gaussian}, $\mathsf{N}_{\gamma}^{\mathsf{full}}$, and the IRLA \eqnref{eqn:GaussianIRLA}, $\mathsf{N}_{\gamma}^{\mathsf{IRLA}}$.
The purple lines correspond to the currently available parameters EuXFEL parameters used in \figref{fig:TotalNumberAngleCurrent}, with the solid line for all signal photons $\mathsf{N}_{\gamma}$ and the dashed line for only photons within the acceptance region of the detector with $\theta_{\gamma} > \theta_{\ast} = 200~\text{\textmu{}rad}$.
The blue dashed line is for the number of photons on the detector acceptance region $\theta_{\gamma} > \theta_{\ast} = 450~\text{\textmu{}rad}$ using the future parameters from the main text.
We find that for the case of including all signal photons the IRLA agrees very well with the results using the full Gaussian pulse in the paraxial approximation, with $\mathsf{E} \lesssim 5\%$ across the full range of the collision angle $\Theta$.
When the detector cut is included, for both currently available and future parameters, we find that for collision angles $\Theta \lesssim 40^{\circ}$ the error becomes quite large, but as the collision angle increases this rapidly falls to the level of $\mathsf{E} \lesssim 1\%$. 
This can be understood by considering what happens when the collision angle $\Theta$ is increased from $\Theta=0$. Initially, the Bragg side peaks are emitted head-on and are excluded by the detector cut. 
As $\Theta$ is increased, the center of the Bragg side peaks move to larger $\theta_{\gamma}$ and eventually to values large enough for their outer edges to fall on the detector. 
To describe the side peak edges accurately, wave-front curvature and the Gouy phase must be included, so for these values of $\Theta$, where the signal is low, the IRLA makes a significant error. 
By experimenting turning on and off various terms in the paraxial beam \eqnref{eqn:Gaussian}, it can be confirmed that the main error is made when both the wavefront curvature and the Gouy terms in the phase are neglected in this cut side peak region.
Once the collision angle $\Theta$ is increased further so that the  Bragg side peak signal is fully on the detector, the error in the IRLA falls to a very low value.
Most importantly, at the collision angle which maximizes $\mathsf{N}_{\gamma,\ast}$ for the currently available parameters, $\Theta = 45^{\circ}$ (vertical dashed lines), and for the future parameters, $\Theta = 63^{\circ}$ (vertical dotted lines), the relative error is on the level of $\mathsf{E} \approx 5\%$ and $\mathsf{E} \approx 0.2\%$, respectively.

\bibliographystyle{apsrev}
\providecommand{\noopsort}[1]{}

\end{document}